\documentclass[authoryear,preprint,review,12pt, a4]{elsarticle}
\usepackage[latin1]{inputenc}
\usepackage{graphics, graphicx}
\usepackage{color}
\usepackage{psboxit,pstcol}
\usepackage{rotating}
\usepackage{amssymb}
\usepackage{lineno}
\usepackage{hyperref}

\biboptions{longnamesfirst}

\journal{Icarus}
\begin{document}
\begin{frontmatter}

\title{The global surface composition of 67P/CG nucleus by Rosetta/VIRTIS. \\
I) Prelanding mission phase}

\author[label1]{Gianrico Filacchione\corref{cor1}} 
\author[label1]{Fabrizio Capaccioni}
\author[label1]{Mauro Ciarniello}
\author[label1]{Andrea Raponi}
\author[label1]{Federico Tosi}
\author[label1]{Maria Cristina De Sanctis}
\author[label2]{St\'ephane Erard}
\author[label2]{Dominique Bockel\'ee Morvan}
\author[label2]{Cedric Leyrat}
\author[label3]{Gabriele Arnold} 
\author[label4]{Bernard Schmitt} 
\author[label4]{Eric Quirico}
\author[label1]{Giuseppe Piccioni} 
\author[label1]{Alessandra Migliorini}
\author[label1]{Maria Teresa Capria}
\author[label1]{Ernesto Palomba}
\author[label1]{Priscilla Cerroni}
\author[label1]{Andrea Longobardo}
\author[label2]{Antonella Barucci}
\author[label2]{Sonia Fornasier}
\author[label5]{Robert W. Carlson}
\author[label3]{Ralf Jaumann}
\author[label3]{Katrin Stephan}
\author[label3]{Lyuba V. Moroz}
\author[label3]{David Kappel}
\author[label2]{Batiste Rousseau}
\author[label6]{Sergio Fonti}
\author[label6]{Francesca Mancarella}
\author[label2]{Daniela Despan}
\author[label4]{Mathilde Faure}

\address[label1]{INAF-IAPS, Istituto di Astrofisica e Planetologia Spaziali, Area di Ricerca di Tor Vergata, via del Fosso del Cavaliere, 100, 00133, Rome, Italy}
\address[label2]{LESIA, Observatoire de Paris, LESIA/CNRS, UPMC, Universit\'e Paris-Diderot, F-92195 Meudon, France}
\address[label3]{Institute for Planetary Research, Deutsches Zentrum f\"ur Luft- und Raumfahrt (DLR), Berlin, Germany}
\address[label4]{Universit\'e Grenoble Alpes, CNRS, IPAG, Grenoble, France}
\address[label5]{Jet Propulsion Laboratory, California Institute of Technology, Pasadena, CA 91109, USA}
\address[label6]{Dipartimento di Matematica e Fisica "Ennio De Giorgi", Università del Salento, Lecce, Italy}

\cortext[cor1]{Corresponding author, email gianrico.filacchione@iaps.inaf.it}

\begin{abstract}
From August to November 2014 the Rosetta orbiter has performed an extensive observation campaign aimed at the characterization of 67P/CG nucleus properties and to the selection of the Philae landing site. The campaign led to the production of a global map of the illuminated portion of 67P/CG nucleus. During this prelanding phase the comet's heliocentric distance decreased from 3.62 to 2.93 AU while Rosetta was orbiting around the nucleus at distances between 100 to 10 km. VIRTIS-M, the Visible and InfraRed Thermal Imaging Spectrometer - Mapping channel \citep{Coradini2007} onboard the orbiter, has acquired 0.25-5.1 $\mu m$ hyperspectral data of the entire illuminated surface, e.g. the north hemisphere and the equatorial regions, with spatial resolution between 2.5 and 25 m/pixel. I/F spectra have been corrected for thermal emission removal in the 3.5-5.1 $\mu m$ range and for surface's photometric response. The resulting reflectance spectra have been used to compute several Cometary Spectral Indicators (CSI): single scattering albedo at 0.55 $\mu m$, 0.5-0.8 $\mu m$ and 1.0-2.5 $\mu m$ spectral slopes, 3.2 $\mu m$ organic material and 2.0 $\mu m$ water ice band parameters (center, depth) with the aim to map their spatial distribution on the surface and to study their temporal variability as the nucleus moved towards the Sun. Indeed,  throughout the investigated period, the nucleus surface shows a significant increase of the single scattering albedo along with a decrease of the 0.5-0.8 and 1.0-2.5 $\mu m$ spectral slopes, indicating a flattening of the reflectance. We attribute the origin of this effect to the partial removal of the dust layer caused by the increased contribution of water sublimation to the gaseous activity as comet crossed the frost-line. The regions more active at the time of these observations, like Hapi in the neck/north pole area, appear brighter, bluer and richer in organic material than the rest of the large and small lobe of the nucleus. The parallel coordinates method \citep{Inselberg1985} has been used to identify associations between average values of the spectral indicators and the properties of the geomorphological units as defined by  \cite{Thomas2015, El-Maarry2015}. Three classes have been identified (smooth/active areas, dust covered areas and depressions), which can be clustered on the basis of the  3.2 $\mu m$ organic material's band depth, while consolidated terrains show a high variability of the spectral properties resulting being distributed across all three classes. These results show how the spectral variability of the nucleus surface is more variegated than the morphological classes and that 67P/CG surface properties are dynamical, changing with the heliocentric distance and with activity processes. 

\end{abstract}

\begin{keyword}
Comets \sep 
Spectroscopy \sep
Composition
\end{keyword}

\end{frontmatter}

\section{Introduction}
Comets are among the most primitive objects in our Solar System. The Jupiter Family Comets, of which 67P/CG is a representative, were formed in the outer region of the early Solar Nebula where temperatures were below 30 K. At these temperatures volatiles, such as H$_2$O, CO$_2$, CO, CH$_3$OH and minor species \citep{Ehrenfreund2004}, and refractory materials, e.g. sulphide pyrrhotite, enstatite and fine-grained porous aggregate material with chondritic composition \citep{Brownlee2006} condense into solid grains, which, along with presolar grains, grew by hierarchical accretion to cometary-sized bodies \citep{MummaCharnley2011} or by gravitationally unstable ``pebble clouds" \citep{Johansen2012}, according to the most recent theories. So far, all cometary nuclei explored by space missions have revealed surfaces mainly covered by dark terrains, and only in some cases, namely comets 9P/Tempel 1 \citep{Sunshine2006, Thomas2013} and 103P/Hartley 2 \citep{AHearn2011, Li2013} few areas showing evidence of exposed water ice-rich units have been observed.

In the low gravity conditions experienced by surface materials on cometary nuclei, the sublimation of volatiles and possibly thermo-mechanical stresses are the major processes responsible for the erosion that shapes the surface and consequently causes degradation, sinkhole collapses \citep{Vincent2015} and mass wasting \citep{Thomas2015}. At heliocentric distances $<$2.7 AU, thermally-driven sublimation of water ice can occur with great efficiency. At the same time mechanical stresses that produce fractures in the upper few meters of the ice matrix can occur due to diurnal thermal expansions and contractions \citep{Lachenbruch1962} or due to pressure build-up caused by volatiles outgassing in inclusions and voids \citep{TauberKurth1987, Auger2015}. 
	
As reported by \cite{Capaccioni2015a}, no evidence of large water ice units on the surface of 67P/CG's nucleus was found in the first data collected by VIRTIS  aboard ESA's Rosetta spacecraft. Global scale data show that the irregularly shaped surface of the nucleus is characterized by morphologically different units \citep{Thomas2015, El-Maarry2015}, which appear uniformly covered by a very dark, dehydrated organic-rich material \citep{Capaccioni2015a}. 
Indeed, the reflectance spectra show two distinct slopes in the VIS and near-IR range as well as a broad absorption band located in the 2.9-3.6 $\mu m$ range. These spectra, as shown by \cite{Capaccioni2015a}, are compatible with a crust  made of a complex mixture of dark  disordered poly-aromatic compounds, opaque minerals and several chemical species containing -COOH, NH$_4^+$, CH$_2$/CH$_3$, -OH (Alcohols) \citep{Quirico2016}. 

\cite{Ciarniello2015} inferred a single scattering albedo of 6.2$\pm$0.2$\%$ at 0.55 $\mu m$ by using VIRTIS data, while \cite{Fornasier2015} reported a value of 6.5$\pm$0.2$\%$ at 0.649 $\mu m$ from OSIRIS observations.
Small high-albedo clusters, about 0.5-1 m wide, isolated or grouped in large clumps up to 30 m across and apparently associated with water ice-rich units, have been recently reported \citep{Pommerol2015, Sierks2015}. 
	
The VIRTIS Spectrometer is described at length in \cite{Coradini2007} and shall not be described again here. The imaging spectrometer VIRTIS-M, operating in the 0.25-5.1 $\mu m$ spectral range with spectral sampling of 1.87 and 9.7 nm/band between 0.25-1.0 and 1.0-5.1 $\mu m$, respectively, has observed water ice in the Hapi active region, in the form of transient deposits formed in the upper layer of the surface by recondensation of water molecules sublimated from sub-surface ice during the diurnal cycle \citep{DeSanctis2015}. In the coma, above the Aten-Babi and Seth-Hapi active regions, VIRTIS-M has observed a strong water vapor emission at 2.67 $\mu m$ \citep{Migliorini2016} while the high spectral resolution spectrometer VIRTIS-H has traced the asymmetric distribution of carbon dioxide and water at northern latitudes, even though $CO_2$ appears more uniform above the comet's nucleus than water vapor \citep{BockeleeMorvan2015}.  Moreover, a clear spatial association between water abundance in the coma and presence of dust has been observed \citep{Capaccioni2015b, Migliorini2016}. The identification of two areas in the Imhotep region during the pre-perihelion passage where exposed water ice is present in stable form at the surface has been reported by \cite{Filacchione2016}.

\section{Observations}
The trajectory of the Rosetta spacecraft around 67P/CG nucleus is quite complex. The low gravity field of the comet allows bound orbits only for distances smaller than 30 km. At higher distances the spacecraft navigates in formation with the nucleus along arcs. Consecutive arcs are interrupted by maneuvers to maintain the spacecraft in the vicinity of the comet. During the pre-landing period the comet's activity - including volatiles outgassing and dust release - was sufficiently low to allow Rosetta to navigate at close distances from the nucleus. In general VIRTIS-M observations are limited by the spacecraft attitude, which is mainly driven by the need to mantain the solar arrays oriented towards the Sun. This implies that the instrument's slit is always oriented perpendicular to the solar direction. The spatial scan occurs therefore always from the solar towards the antisolar direction.
VIRTIS observations are scheduled according to the mission timelines, subdivided into Long Term Plan (LTP, four months before execution), MTP (Medium Term Plan, one month before execution), STP (Short Term Plan, one week before execution) and VSTP (Very Short Term Plan, 3.5 days before execution).

\begin{figure}[h!]
	\centering
		\includegraphics[width=16cm]{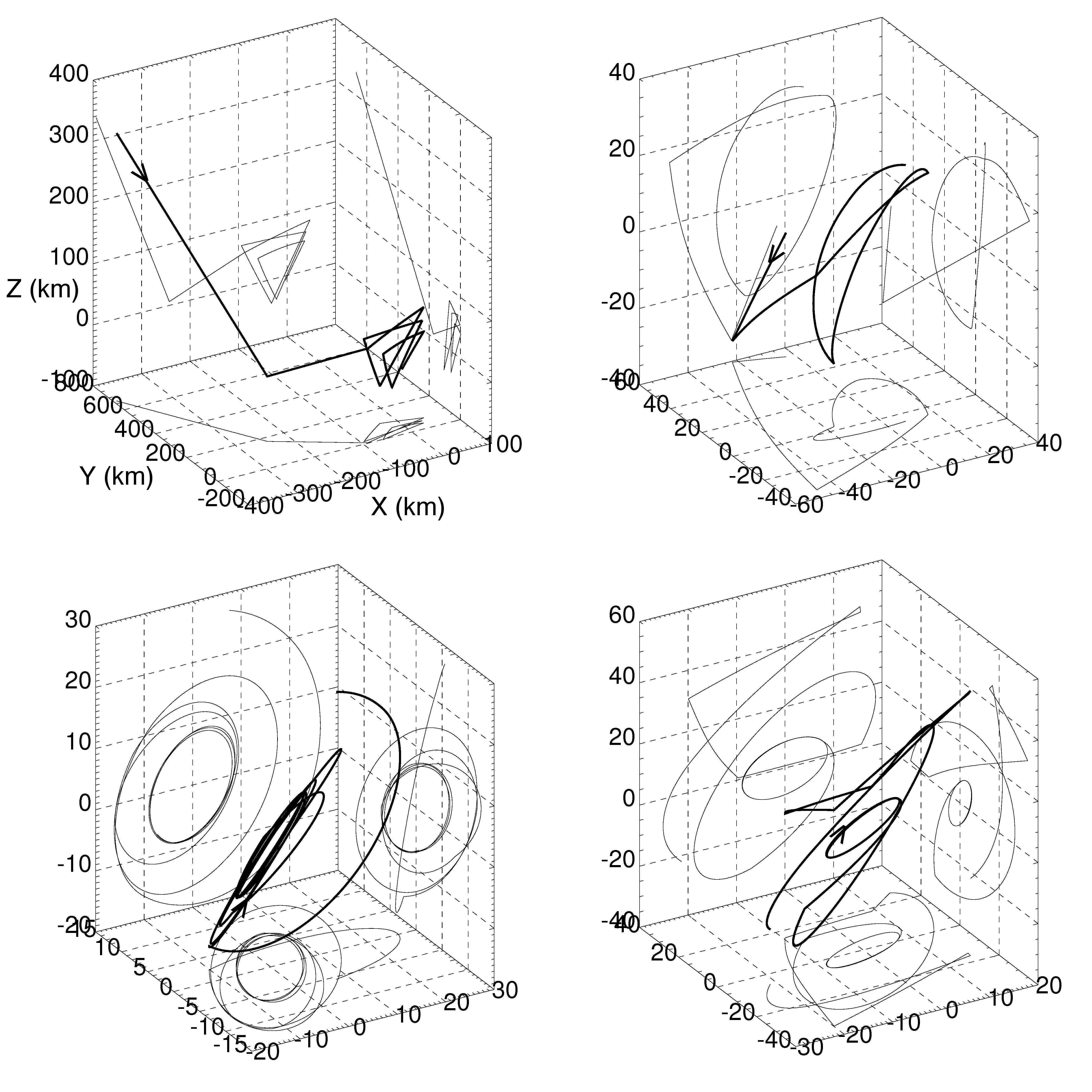}
		\caption{Rosetta's trajectory around comet nucleus (bold lines) during MTP6 (top left panel), MTP7 (top right), MTP8 (bottom left), MTP9 (bottom right). In each panel the nucleus center is at the origin of the axes. The black arrow indicates the starting point for each trajectory. Projections on planes XY, XZ and YZ are also shown (thin lines).}
		\label{fig:figure1}
\end{figure}

\begin{figure}[h!]
	\centering
		\includegraphics[width=14cm]{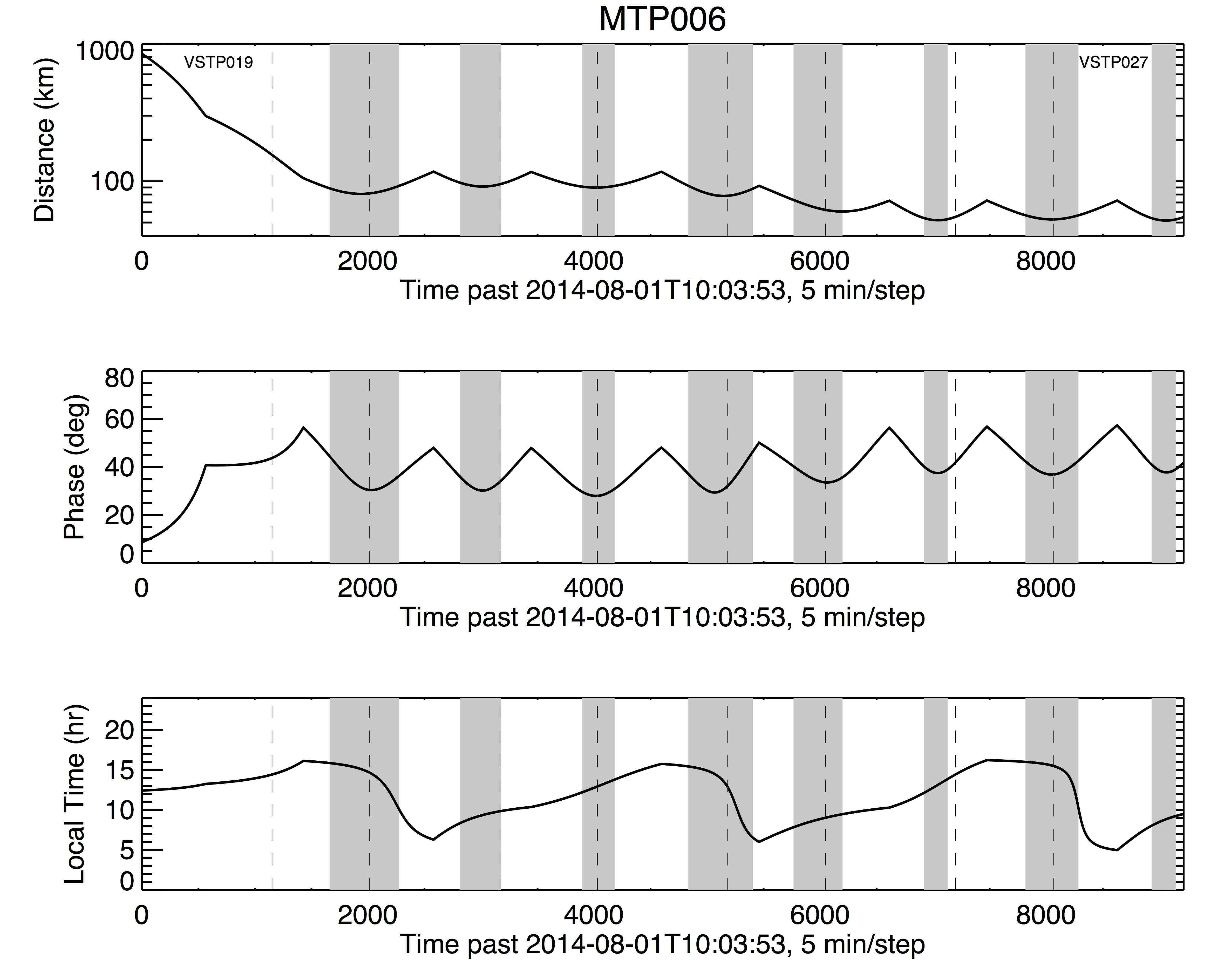}
		\caption{VIRTIS-M observations timeline during MTP6 mission period shown as a function of Rosetta-comet distance (top panel), solar phase (central panel) and local time (bottom panel). These parameters refers to the nadir point on the nucleus surface. Time axis values, expressed at 5 minutes increments, are counted from 2014-08-01T10:03:53 corresponding to the MTP6 starting time. Vertical dashed lines separate VSTP mission phases. Gray boxes indicate the timing of VIRTIS-M mapping sequences ID from 1 to 8 as listed in Table \ref{table:2}.} 
	\label{fig:figure2}
\end{figure}

\begin{figure}[h!]
	\centering
		\includegraphics[width=14cm]{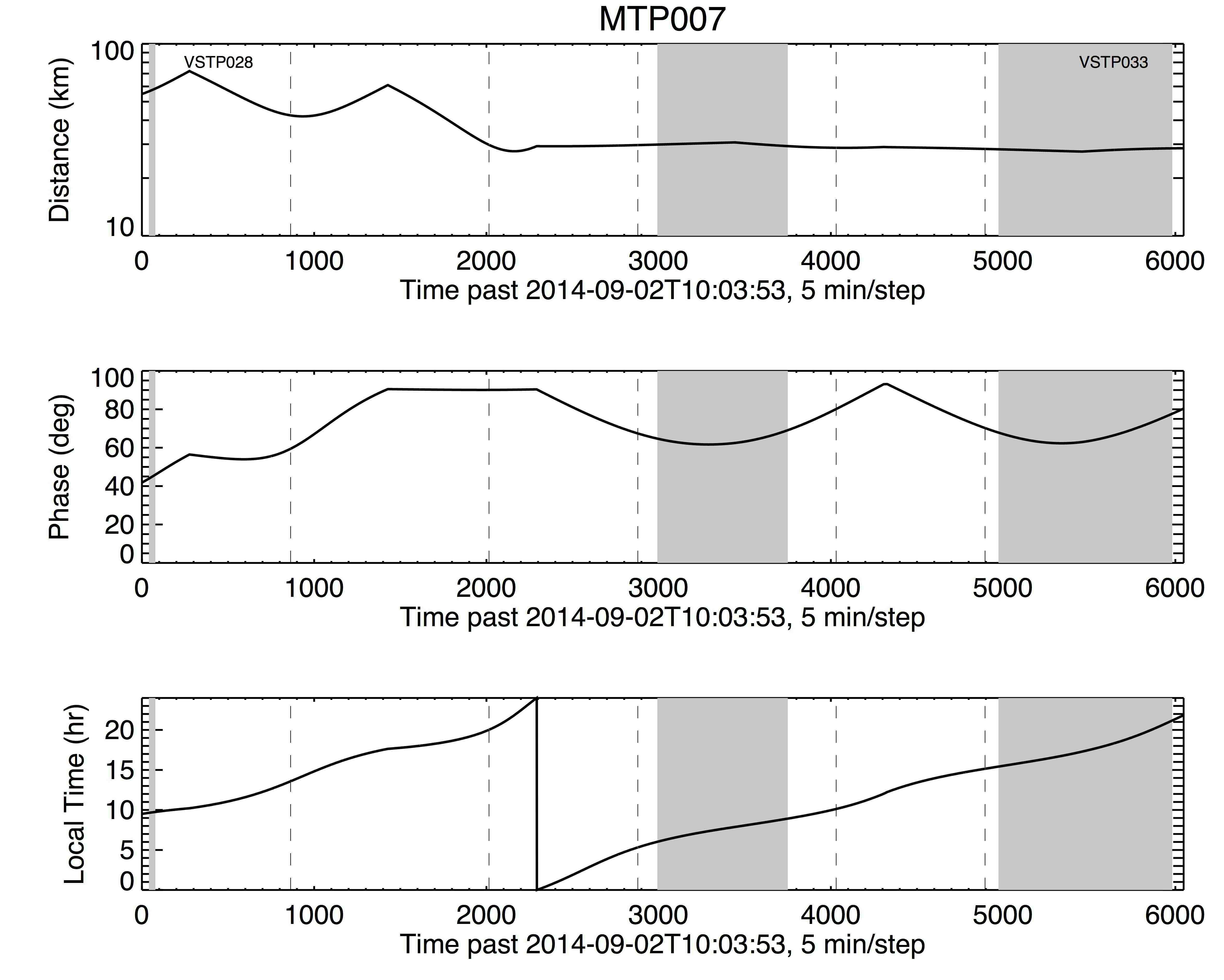}
		\caption{VIRTIS-M observations timeline during MTP7 mission period starting at 2014-09-02T10:03:53. Same scheme as Fig. \ref{fig:figure2}. Gray boxes indicate the timing of VIRTIS-M mapping sequences ID from 9 to 11 as listed in Table \ref{table:2}.}
		\label{fig:figure3}
\end{figure}

\begin{figure}[h!]
	\centering
		\includegraphics[width=14cm]{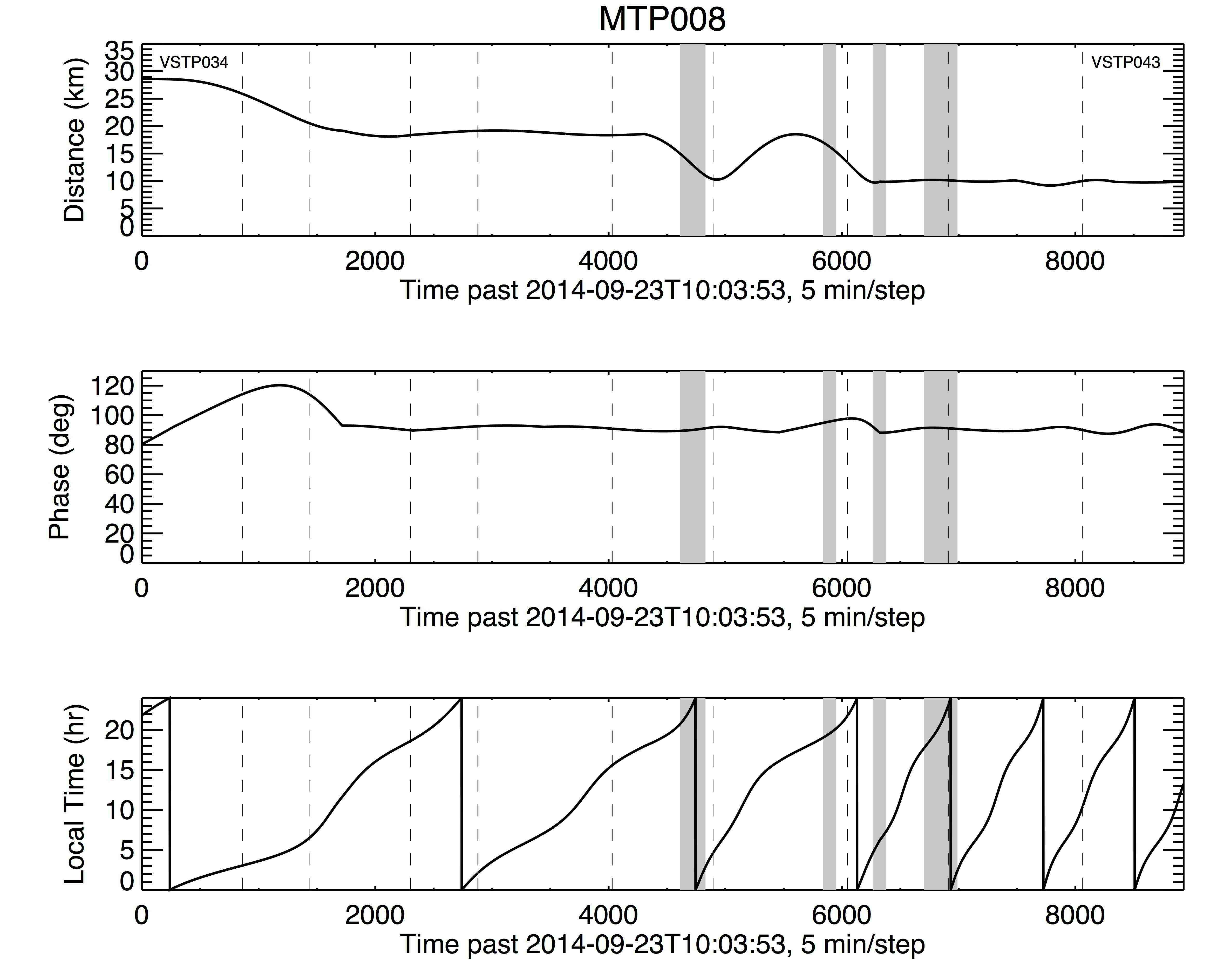}
		\caption{VIRTIS-M observations timeline during MTP8 mission period starting at 2014-09-23T10:03:53. Same scheme as Fig. \ref{fig:figure2}. Gray boxes indicate the timing of VIRTIS-M mapping sequences ID from 12 to 15 as listed in Table \ref{table:2}.}
		\label{fig:figure4}
\end{figure}

\begin{figure}[h!]
	\centering
		\includegraphics[width=14cm]{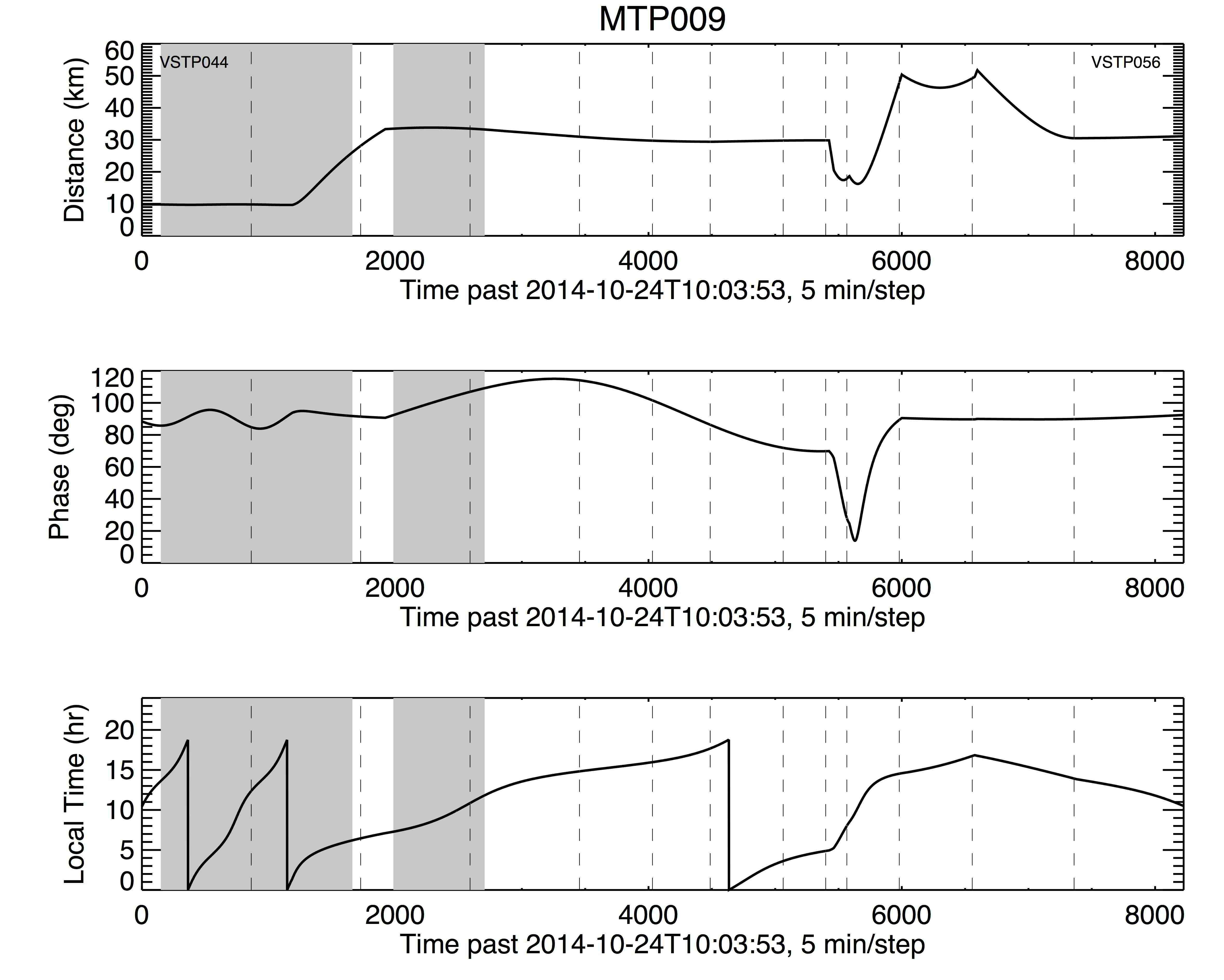}
		\caption{VIRTIS-M observations timeline during MTP9 mission period starting at 2014-10-24T10:03:53. Same scheme as Fig. \ref{fig:figure2}. Gray boxes indicate the timing of VIRTIS-M mapping sequences ID 16-17 as listed in Table \ref{table:2}.}
		\label{fig:figure5}
\end{figure}

During the Philae pre-landing mission period, VIRTIS-M has completed four different nucleus mapping campaigns: the first in August 2014, during the MTP6 mission period, when the spacecraft was on pyramid orbits at a distance of 50-100 km from the nucleus with 25$^\circ$-40$^\circ$ solar phase angle. During this MTP a large number of observations have been executed at constant solar phase (about 30$^\circ$) with spatial resolution of 12.5 to 25 m/pixel. The second phase occurred in September 2014, during MTP7, when the spacecraft was on petals or arcs orbits at about 30 km distance with 60$^\circ$-70$^\circ$ solar phase angle. During this period VIRTIS-M has observed the nucleus in oblique view with a spatial resolution of about 7.5 m/pixel. The third campaign (MTP8) was executed in October 2014 when the spacecraft was on terminator circular orbits at 20 and then 10 km distances, resulting in observations with a spatial resolution of 5 to 2.5 m/pixel taken at solar phase angle of 90$^\circ$.  Finally, the last campaign (MTP9) was executed from the end of October to mid-November 2014, in preparation of Philae lander release when the spacecraft's trajectory was on elliptic orbits between 10 and 20 km from the nucleus, with VIRTIS-M continuing to monitor the surface with spatial resolution between 2.5 and 5 m/pixel with solar phase angle between 70$^\circ$ to 100$^\circ$.       
A summary of the mission periods timelines is given in Table \ref{table:1}. The Rosetta trajectory around the comet nucleus is shown in Fig. \ref{fig:figure1}. The VIRTIS observations timeline is listed in Table \ref{table:2} and shown in Fig. \ref{fig:figure2} - \ref{fig:figure5} as function of the Rosetta-nucleus distance, solar phase angle and local time for MTP6, 7, 8, 9, respectively.

\begin{table}
\centering  
\begin{tabular}{|c|c|l|}      
\hline\hline
MTP & Date & Orbit description \\
\hline
MTP6 &  2014-08-06 / 2014-08-17 & First Pyramid Orbit averaging 100 km distance \\
     &  2014-08-07              & Arrival, begin first leg \\
     &  2014-08-10              & Begin second leg \\     
     &  2014-08-13              & Begin third leg \\
     &  2014-08-17 / 2014-08-24 & Transfer to 50 km distance\\
     &  2014-08-20              & Return at 80 km distance\\
     &  2014-08-24 / 2014-09-03 & Second Pyramid Orbit averaging 50 km distance \\
     &  2014-08-24              & Begin first leg \\
     &  2014-08-27              & Begin second leg \\
     &  2014-08-31              & Begin third leg \\     
\hline
MTP7 & 2014-09-03 / 2014-09-10  & Petal orbits at 30 km distance, afternoon LST \\
     & 2014-09-10 / 2014-09-23  & Petal orbits at 30 km distance, morning LST \\
\hline
MTP8 & 2014-09-23 / 2014-09-28  & Night excursion and transfer to 20-km bound orbit \\
     & 2014-09-28 / 2014-10-08  & Terminator bound orbit at 20 km \\
     & 2014-10-08 / 2014-10-24  & Terminator bound orbit at 10 km \\     
\hline
MTP9 & 2014-10-24 / 2014-10-28  & Terminator bound orbit at 10 km \\
     & 2014-11-11               & Philae lands \\
\hline \hline
\end{tabular}
\caption{Summary of Rosetta's orbits during Medium Term Plan MTP6-MTP9 periods.}  
\label{table:1}                   
\end{table}

\begin{table}
\centering  
\begin{tabular}{|c|c|c|c|c|}      
\hline\hline
MTP Period & Mapping Sequence ID &  Start Time & End Time & Nr Observations \\
\hline \hline
6 & 1 & 2014-08-07T04:19:48 & 2014-08-09T08:55:02 & 44 \\
6 & 2 & 2014-08-11T04:19:47 & 2014-08-12T09:40:02 & 20 \\
6 & 3 & 2014-08-14T23:19:45 & 2014-08-15T23:54:56 & 21 \\ %
6 & 4 & 2014-08-18T04:19:42 & 2014-08-20T03:54:54 & 32 \\
6 & 5 & 2014-08-21T11:19:44 & 2014-08-22T21:54:52 & 24 \\
6 & 6 & 2014-08-25T11:04:19 & 2014-08-26T05:54:52 & 16 \\
6 & 7 & 2014-08-28T14:29:51 & 2014-08-30T03:54:57 & 7 \\
6 & 8 & 2014-09-01T11:04:43 & 2014-09-02T04:45:57 & 7 \\
\hline
7 & 9 & 2014-09-02T12:00:30 & 2014-09-02T17:50:24 & 3 \\
7 & 10 & 2014-09-12T18:34:51 & 2014-09-15T09:51:04 & 28 \\
7 & 11 & 2014-09-19T15:09:51 & 2014-09-23T04:30:51 & 16 \\
\hline
8 & 12 & 2014-10-09T11:15:22 & 2014-10-10T05:43:00 & 7 \\
8 & 13 & 2014-10-13T15:15:20 & 2014-10-14T00:28:20 & 9 \\
8 & 14 & 2014-10-15T04:09:40 & 2014-10-15T13:25:31 & 11 \\
8 & 15 & 2014-10-16T16:12:52 & 2014-10-17T17:45:24 & 15 \\
\hline
9 & 16 & 2014-10-24T23:14:50 & 2014-10-30T04:40:24 & 66 \\
9 & 17 & 2014-10-31T07:39:45 & 2014-11-02T20:01:52 & 36 \\
\hline \hline
\end{tabular}
\caption{VIRTIS-M observations summary for MTP mission phase. For each mapping sequence are reported the time interval and the number of acquired observations.}  
\label{table:2}                   
\end{table}

Given the nucleus irregular bilobate shape \citep{Sierks2015} and the orientation of the spin axis \citep{Preusker2015}, the north hemisphere is mainly illuminated at aphelion (5.68 AU) when the sub-solar latitude is at +49$^\circ$. Moving inwards to the Sun the maximum subsolar latitude of +52$^\circ$ is reached before zeroing at the vernal equinox occurring in early May 2015 at a heliocentric distance of 1.71 AU. Around perihelion, occurring in mid August 2015 at 1.24 AU, mainly the south hemisphere is illuminated. In this period the subsolar latitude reaches the southernmost point, corresponding to a latitude of -52$^\circ$ about 20 days post-perihelion. Moving outwards from the Sun, the autumnal equinox is reached at beginning of April 2016 at 2.69 AU, when the sub-solar latitude is again above the equator. Therefore in studying the 67P/CG nucleus we need to take into account these orbital characteristics which strongly influence the intensity and distribution of the solar flux received by the comet: the seasonal excursion of the subsolar point is such that while the northern regions are illuminated by the Sun for a long period of time when the comet is at aphelion, the southern regions collect a much more intense solar flux during the rapid perihelion passage. In this paper we are analyzing northern and equatorial region data collected by VIRTIS-M in a period of time spanning from August to November 2014, when the comet orbital position was at a heliocentric distance reducing from 3.62 to 2.93 AU. 

\section{Data processing}
\subsection{Calibration and a-posteriori corrections}
VIRTIS-M raw data are converted to spectral I/F by applying the calibration pipeline described in \cite{Filacchione2006, Ammannito2006, Filacchione_etal_2006} with the additional corrections derived from in-fight data \citep{Migliorini2013, Raponi2014}. The pipeline includes the removal of detector's dark current and background noise, as measured by the instrument by closing the spectrometer's slit shutter at periodic time intervals. For the observations used in this work, the shutter was commanded in close position at the beginning of each cube and then every 20 frames until the end of the observation. Dark and background signals are interpolated for this purpose to the time of each science acquisition (frame) in order to remove possible drifts. 
While the Dark current contribution is present throughout the full spectral range and is only sensititve to IR detector temperature variations, the background signal is relevant in the thermal spectral range, between 3.5 and 5 $\mu m$, and is directly correlated to the thermal flux emitted by the spectrometer's enclosure which can change during long observation sequences. The nominal temperature for the spectrometer's enclosure is 135 K, but depending on the environmental conditions this temperature can increase up to about 140 K.
Moreover, saturated and defective pixels are flagged and removed from scientific analysis. Finally, the visible channel is further corrected for the spectral tilt effect, a parallelogram distortion of the signal on the focal plane caused by a non perfect coalignment among spectrometer's slit, visible grating grooves direction and CCD detector: the consequences of this effect in the data and the method applied to remove it are described in \cite{Filacchione2006}.

\begin{figure}[h!]
	\centering
		\includegraphics[width=14cm]{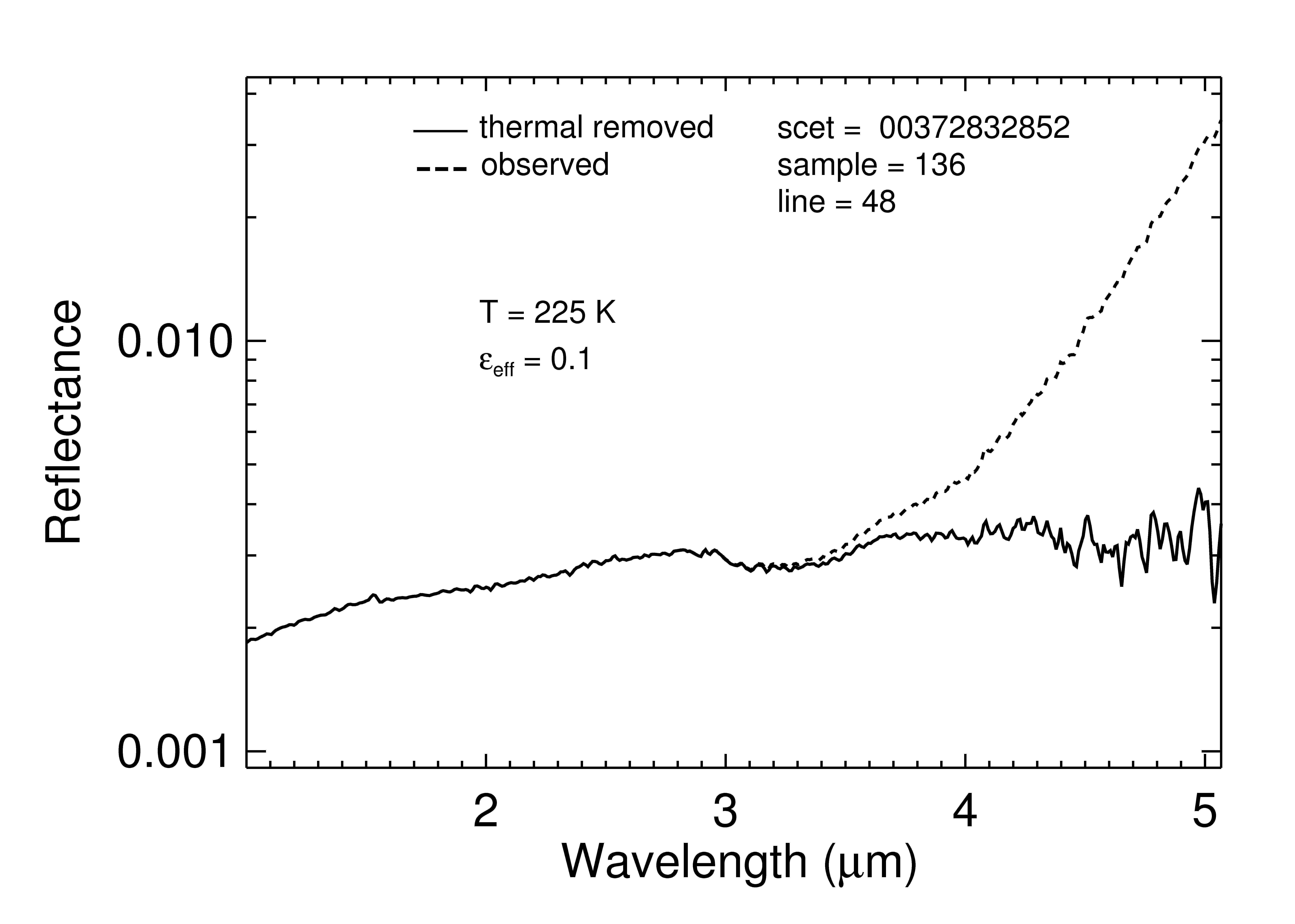}
		\caption{Example showing the thermal emission removal to obtain reflectance spectra in the 1-5 $\mu m$ spectral range.}
		\label{fig:figure6}
\end{figure}

In order to properly study the composition of the nucleus, additional postprocessing reduction has been performed to derive spectral reflectance from I/F: the photometric modelling, necessary to remove illumination effects from the data, and the removal of the nucleus thermal emission in the 3.5-5 $\mu m$ range. Details about photometric modelling are given in \cite{Ciarniello2015} and shall be not repeated here. As shown in Fig. \ref{fig:figure6} the nucleus thermal emission contribution become extremely relevant in the 3.0-5 $\mu m$ range as the nucleus gets closer to the Sun and its surface temperature increases. In this conditions the thermal emission affects the interpretation of the spectrum longward of 3.0 $\mu m$ (and notably also of the 3.2 $\mu m$ organic band) and has to be removed. The radiance spectra are modeled as the sum of the solar radiance reflected by the nucleus surface and of the nucleus thermal emission flux. Thermal emission is removed by modeling radiance spectra as the sum of the reflected flux and the thermal emission. A similar approach was used to model Deep Impact data of comet 9P/Tempel 1 \citep{Sunshine2007}.
 The reflected component is calculated for each spectrum from the parameters derived by means of the photometric modeling \citep{Ciarniello2015}. In particular, single scattering albedo in the range 3.5 - 5 $\mu m$ is extrapolated from shorter wavelengths as a continuum with a fixed slope. This approach introduces some arbitrarity  in the spectral slope of the retrieved spectrum. However, it does not alter the capability to retrieve any spectral feature should they be present.
The reflectance calculated in this way is then multiplied by a free parameter to account for the measured level of signal. The thermal emission is then modeled as a gray body with two free parameters: temperature (T) and emissivity ($\epsilon_{eff}$). The free parameters are retrieved by a Levenberg-Marquardt least squares optimization algorithm, then the modeled thermal emission is subtracted from the measured spectra. Due to the spurious signal generated by the instrument internal background (as described previously) the signal-to-noise ratio in the thermal range can be considerably lower than at shorter wavelengths. 

The adopted methodology, due to the relatively low signal to noise ratio in this spectral range, is more sensitive to the presence of calibration residual in the instrument responsivity which will show up as spurious absorption and/or emission features in the resulting reflectance spectrum. For this reason we do not attempt to interpret the  spectral characteristics which are observed in the 3.5-5 $\mu m$ range, see Fig. \ref{fig:figure6}. For some observations, for instance those taken in the central hours of the cometary day and affected by the largest thermal emission, also the long wavelengths portion of the organic feature around 3.2 $\mu m$ is affected by the presence of thermal emission which alter both the local continuum and the band center. These observations have been discarded from the present analysis.

\subsection{Data projection and mapping}
All the geometric parameters (latitude, longitude, incidence angle, emission angle, phase angle, distance and local solar time) required to derive the true reflectance of the surface, have been computed for each VIRTIS-M pixel intercepting the nucleus surface. These parameters are computed for the pixel's center and for the four corners by means of SPICE routines \citep{Acton1996} and the spacecraft's trajectory and attitude kernels. The nucleus shape model used for computation and data projection is SHAP5 v1.1, derived from OSIRIS images by using a stereophotoclinometry method \citep{Jorda2014}. Three further effects are considered in the data projections: the first is the presence of large portions of the dayside surface affected by shadows, in general occurring above Hapi, Babi and Seth regions. The small lobe's shadows are cast over these regions during the time interval considered in this work. In general all pixels falling on points of the surface not visible from the Sun are filtered out before mapping. Furthermore, only pixels where the distances of the four corners from the pixel center are less than 5$^\circ$ are processed: this filtering is necessary to exclude pixels falling in between the two lobes taken during oblique views of the nucleus and therefore  spatially inconsistent. Finally, pixels illuminated or observed in very oblique geometries, e.g. having incidence $i \ge 80^\circ$ or emission $e \ge 80^\circ$ angles with respect to the local normal direction, have been filtered out before producing the final maps.

\begin{figure}[h!]
	\centering
		\includegraphics[width=17cm]{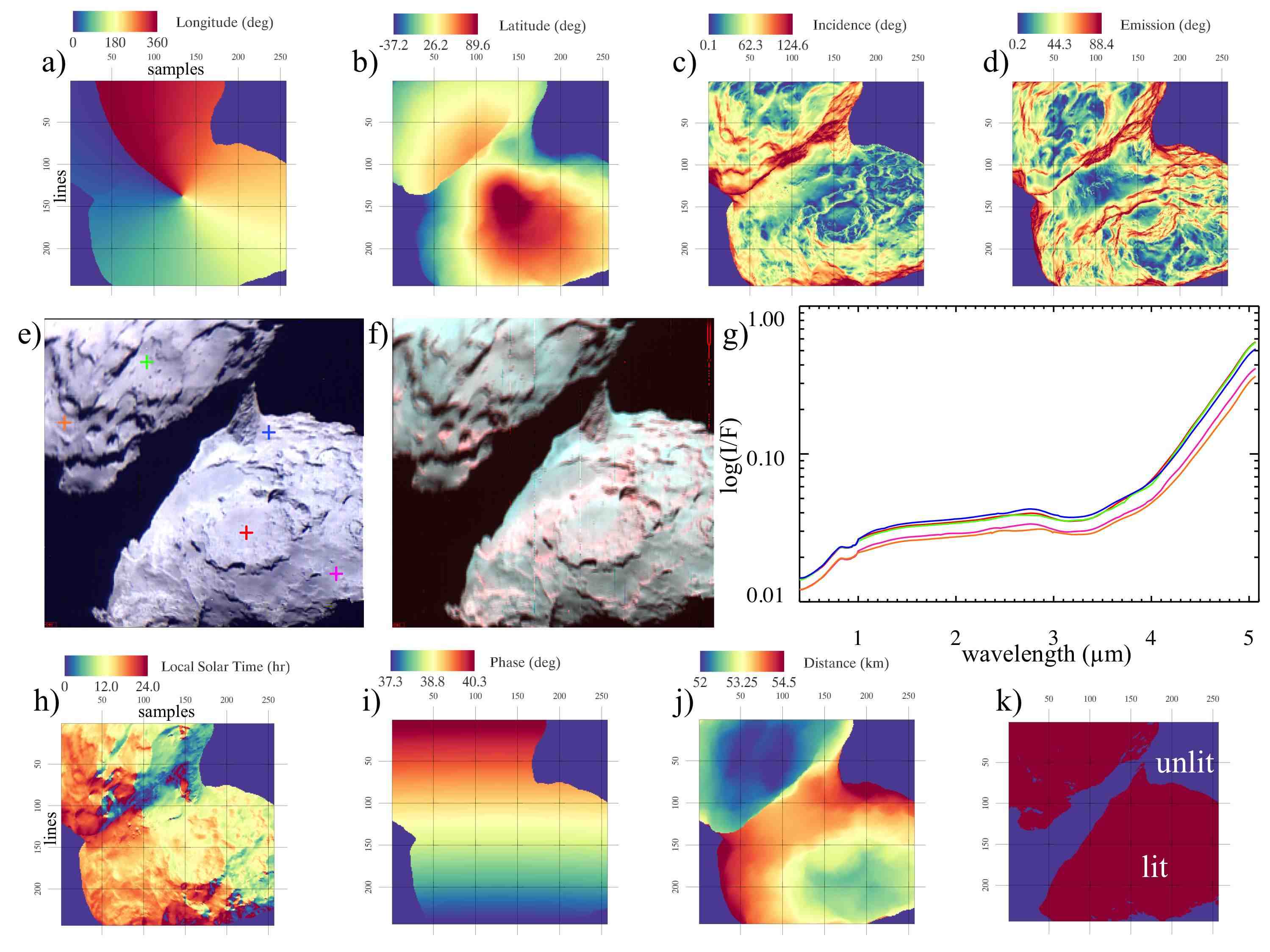}
		\caption{Hyperspectral data and associated geometry parameters for a typical nucleus observation (cubes V1$\_$00367957721.QUB, I1$\_$00367957714.QUB). \emph{Top row:} image frame of longitude (panel a), latitude (panel b), incidence angle (panel c), emission angle (panel d). \emph{Center row:} VIS channel I/F color image (B=0.44 $\mu m$, G=0.55 $\mu m$, R=0.7 $\mu m$, panel e). The colored crosses identify five 5$\times$5 pixels regions for which we report the average I/F VIS-IR spectra in panel g); IR channel I/F color image (B=1.5 $\mu m$, G=3 $\mu m$, R=4.5 $\mu m$, panel f); average I/F spectra for the five regions (panel g). Data are not corrected for photometry response and thermal emission is not removed. \emph{Bottom row:} image frame of local solar time (panel h), phase angle (panel i), distance from spacecraft (panel j), lit/unlit (maroon/blue color code) pixels image mask (panel k). The spatial resolution of this image is about 13 m/pixel.} 
		\label{fig:figure7}
\end{figure}

A typical VIRTIS-M nucleus observation correlated with the geometry parameters computed for the pixel's center is shown in Fig. \ref{fig:figure7}.
This particular acquisition was returned by VIRTIS-M on August 29th 2014 between 18:29:50 and 19:55:04 UTC time (MTP6 mission phase) and corresponds to a high resolution 256$\times$256 samples by lines, 864 bands cube. Each line along the vertical axis of the figure was acquired every 20 seconds by moving the internal scan mirror. During each line step, the VIS and IR detectors were acquiring with integration times of 10 and 3 seconds, respectively. By adopting the instrument's samples-lines reference system, the spectral information associated to each pixel (I/F, spectral reflectance and derived spectral indicators as discussed ahead) can be directly linked to the geometry quantities.          

A cylindrical projection with a regular grid of 0.5$^\circ \times$0.5$^\circ$ resolution both in longitude and latitude is used to map and render VIRTIS-M data. Each bin corresponds to a linear distance of about 15$\times$15 m. A cylindrical map offers the advantage to simplify the data projection process but has the drawback to not guarantee the rectification of the points on the map (distances among points are non-uniform). Moreover, the irregular shape of the nucleus and the location of the north pole on the large lobe in the Hapi region causes a strong degeneration of the coordinates grid in the regions nearby the neck, where up to three distinct points on the surface can have the same set of longitude-latitude values. 

\begin{figure}[h!]
	\centering
		\includegraphics[width=17cm]{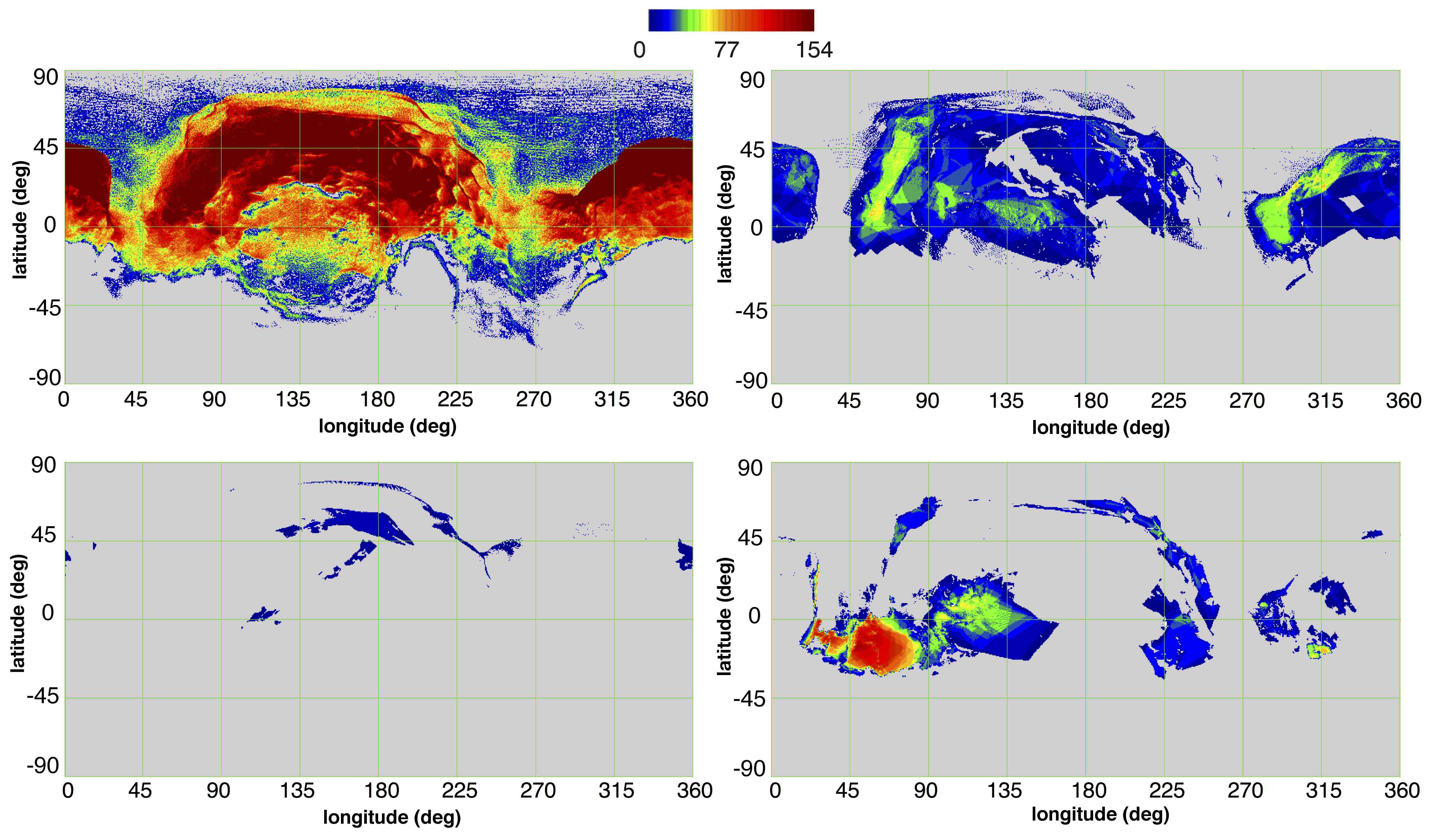}
		\caption{Data redundancy (number of observations per bin) cylindrical maps for MTP6 (top left), MTP7 (top right), MTP8 (bottom left), MTP9 (bottom right).}
		\label{fig:figure8}
\end{figure}

Since VIRTIS-M dataset offers a wide data redundancy, with up to 154 pixels intercepting a given 0.5$^\circ \times$0.5$^\circ$ bin on the map during the MTP6 period, we have applied the following method to reduce projected data grouped together for MTP6 to MTP9 phases: 1) each single pixel has been calibrated, despiked, corrected for photometry response and for thermal emission as previously described. Geometry parameters, e.g. incidence, emission, phase angle and local solar time, have been calculated with respect to the local facet as derived from the shape model; 2) independently from local solar time, each single valid pixel has been projected on the grid by considering the position of the center and four corners;  3) for a given bin, the median value is shown in the maps discussed thereafter. The redundancy maps are shown in Fig. \ref{fig:figure8}: in the pre-landing phase the best coverage is achieved during the MTP6 period when a contiguous coverage across the north hemisphere and equatorial region was achieved. For Atum region the maximum meridional coverage is reached at latitude = -65$^\circ$  while the Imhotep plain is completely covered. On many regions of the body and of the small lobe, the redundancy value is up to 154 observations/bin. For reference, the identification of the morphological regions, as defined by \cite{Thomas2015, El-Maarry2015}, is shown in Fig. \ref{fig:figure10}. A more sparse coverage is obtained during the MTP7 period where Hapi, Atum, Anuket and Anubis regions were not observed. In this period a maximum redundancy of 60 observations/bin is achieved above Khepry-Babi. MTP8 and MTP9 have scattered distributions, with more contiguous points observed mainly on Seth-Ash on the body and Ma'at on the small lobe. MTP9 offers also a good coverage for Imhotep, Kephry and Aker where the redundancy is about 100 observations/bin.   
 
\section{CSI: Comet's Spectral Indicators}
Since the nucleus of 67P/CG appears remarkably uniform in the VIS-IR spectral range, showing a very low geometric albedo equal to $A_{geo}$(0.55 $\mu m$) = 6.2 $\pm$ 0.2$\%$ \citep{Ciarniello2015}, we have adopted a set of Comet's Spectral Indicators (CSI) able to trace composition and physical properties of the surface. These indicators are optimized to study the typical 67P/CG nucleus reflectance spectra as shown in Fig. \ref{fig:figure7} - panel g).  
The region between 0.85-1.0 $\mu m$ displays a spurious absorption feature most probably attributable to calibration residual and low signal to noise conditions. For this reason we have not used this spectral interval to derive the slope in the VIS range.

From the average spectral properties described in section 3, the study of the nucleus colors and composition is carried out by means of the following CSI (Comet's Spectral Indicators):
\begin{itemize}
\item SSA, the Single Scattering Albedo, computed after converting I/F data to spectral reflectance by applying photometric correction and removing thermal emission;
\item $S_{0.5-0.8 \mu m}$, the visible spectral slope, corresponding to the average slope computed by a linear best-fit of the reflectance in the 0.5-0.8 $\mu m$ range; the I/F is normalized at 0.55 $\mu m$ before computing the fit in order to remove residual illumination effects and to decouple the color variability from apparent brightness (mainly due to the solar phase changes).
\item $S_{1.0-2.5  \mu m}$, the infrared spectral slope, corresponding to the average slope computed by best-fitting the reflectance in the 1.0-2.5 $\mu m$ range with a line; it is computed by applying the same normalization at 0.55 $\mu m$ as the visible spectral slope. 
\item BD(3.2 $\mu m$) and BC(3.2 $\mu m$), the 3.2 $\mu m$ organic material absorption band depth and center, respectively.
\item BD(2.05 $\mu m$), the water ice 2.05 $\mu m$ absorption band depth.
\end{itemize}
Band depths and centers \citep{Clark1999} are computed after having normalized the bands' reflectance with respect to the local continuum. Band depth is calculated on the wavelength of the band center after having removed the continuum. 
The continuum is computed as a straight line by fitting reflectance data between 2.69-2.71 and 3.71-3.73 $\mu m$ for the 3.2 $\mu m$ band and between 1.80-1.82 and 2.22-2.24 $\mu m$ for the 2.05 $\mu m$ band, respectively.

The method to compute spectral slopes is given in \cite{Filacchione2010, Filacchione2012}. In the next paragraphs we discuss the resulting cylindrical maps rendered for each one of these spectral indicators.    

\subsection{Single Scattering Albedo SSA cylindrical map}

\begin{figure}[h!]
	\centering
		\includegraphics[width=17cm]{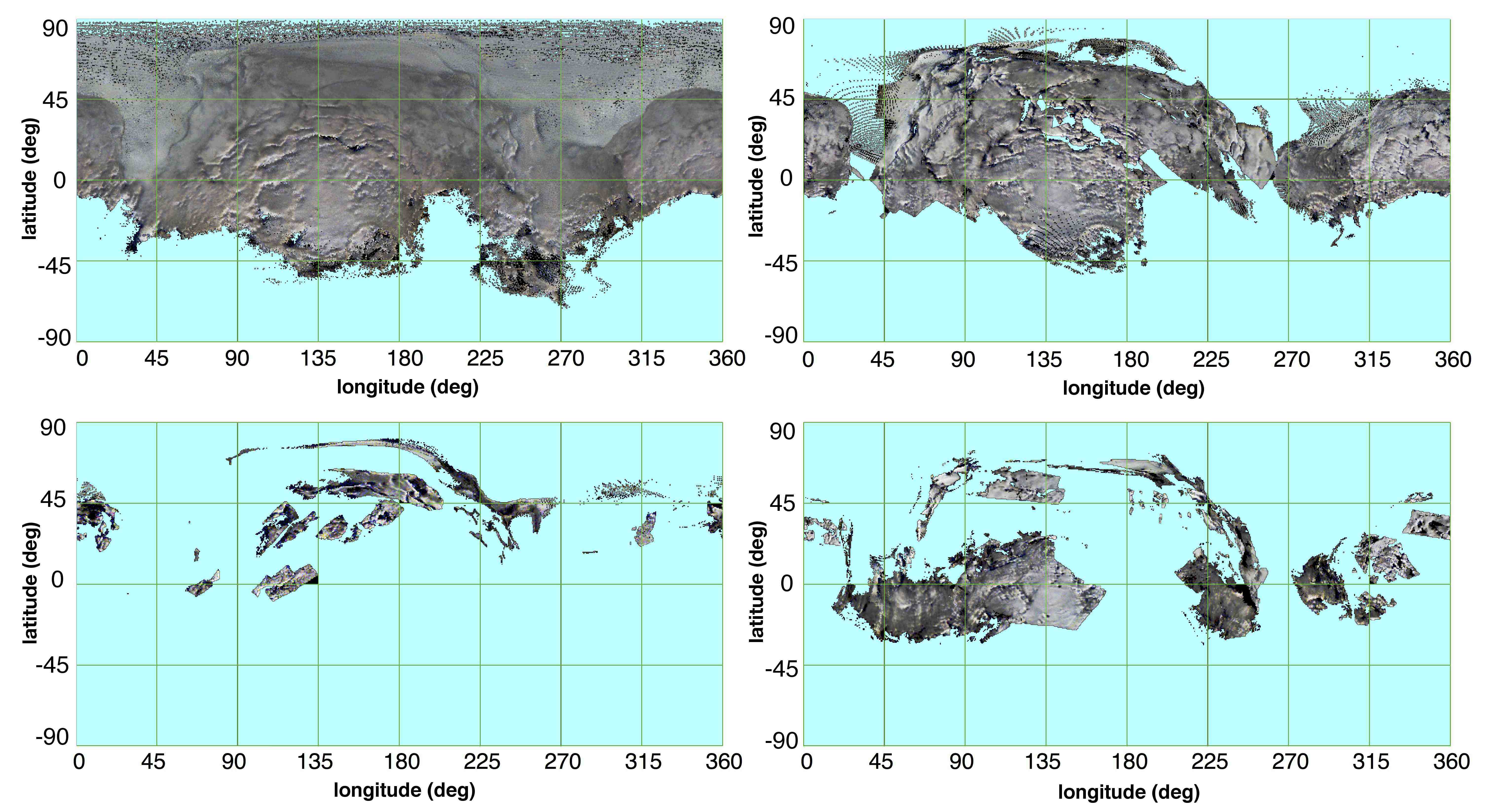}
		\caption{Visible color single scattering albedo cylindrical maps for MTP6 (top left), MTP7 (top right), MTP8 (bottom left), MTP9 (bottom right). Each map is composed by blue B=0.44 $\mu m$, green G=0.55 $\mu m$ and red R=0.7 $\mu m$ channels.}
		\label{fig:figure9}
\end{figure}

Surface composition, regolith grain size and presence of condensed volatiles control the values of the SSA across the different regions of the comet resulting in color changes. We are adopting a scheme in which a combination of three visible SSAs (blue B=0.44 $\mu m$, green G=0.55 $\mu m$ and red R=0.7 $\mu m$) is used to render nucleus colors as shown in Fig. \ref{fig:figure9}. On average, the brightest regions of the nucleus are Serget, Hapi and Ma'at. Several units of the neck/north pole in the Hapi region and in the Imhotep central plain show high SSA values, appearing locally brighter than the rest of the nucleus. Higher SSA values are certainly correlated with the presence of water ice in the outer layers. Water ice has been in fact identified by VIRTIS-M on debris falls in the Imhotep region where it is exposed on the walls of elevated structures, below overhangs and accumulating in the debris fields at the base of these features \citep{Filacchione2016}. In the Hapi region VIRTIS-M has recognized and followed the diurnal cycle of water ice sublimating and recondensing in phase with illumination changes occurring in September 2014 \citep{DeSanctis2015}. Moreover, in April 2015 VIRTIS-M has observed the maximum H$_2$O emission above Aten-Babi and Seth-Hapi regions \citep{Migliorini2016}, a result indicating the presence of surface water ice and activity occurring in those areas. The darkest terrain units are identified across Ash, Aker and Serget, in agreement with OSIRIS spectrophotometry \citep{Fornasier2015}. 

\begin{figure}[h!]
	\centering
		\includegraphics[width=17cm]{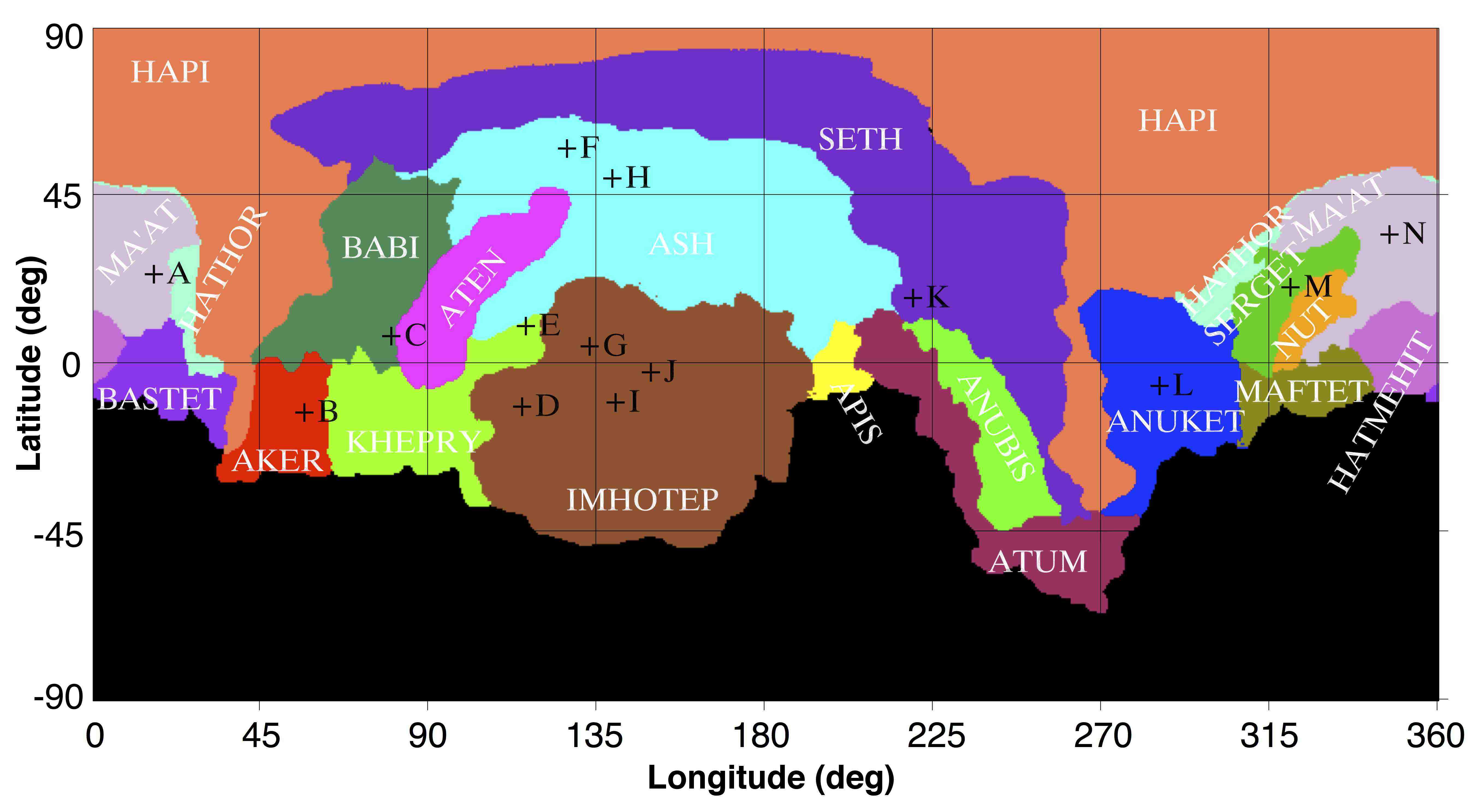}
		\caption{Cyclindrical map of 67P/CG's nineteen geomorphological regions (white labels) as defined by \cite{Thomas2015, El-Maarry2015}. The position of the fourteen GCPs labelled from A to N (black labels) are marked. GCP spectral indicator values are reported in Tables \ref{table:3} - \ref{table:4}.} 
		\label{fig:figure10}
\end{figure}

The availability of maps taken at different heliocentric distances allows us also to study the temporal variability of the SSA across the surface. VIRTIS-M data clearly show that with the comet moving towards perihelion a general increase of the SSA values on wide areas located on both bright and dark terrains is occurring: this effect is probably caused by the progressive removal of the dust layer caused by gaseous activity with the consequent exposure of more fresh and bright material. 
At the same time water ice is also sublimating and disappearing from the surface because the nucleus is moving towards the Sun. In the time period analyzed in this work the removal of the dust is much more efficient than the water ice sublimation, resulting in an increase of the apparent brightness of the nucleus. This trend continues to be observed in the data taken during the successive Rosetta escort phase (from November 2014 to May 2015) which will be discussed in the next paper II). Modeling the dust to volatiles ice fraction on the surface is a task beyond the scope of our analysis.
This surface brightening is well-evidenced by the SSA(0.55 $\mu m$) values we report in Table \ref{table:3} for 14 Ground Control Points (GCP), each 0.5$^\circ \times$0.5$^\circ$ wide, taken across the surface during MTP6-7-9 periods when the heliocentric distance was decreasing from 3.62 to 3.31 and finally to 2.93 AU, respectively. The investigated points are selected on the basis of their observability and coverage occurring during these three periods. The positions of the GCP are indicated by letters A to N in Fig. \ref{fig:figure10} on a cylindrical map showing also the extension of the geomorphological regions on the nucleus as defined by \cite{Thomas2015} and \cite{El-Maarry2015}. MTP8 data are not used in this analysis because their spatial coverage is very sparse and patchy. A significant increase of the SSA is occurring on each of these GCPs, with a maximum increase of about 87$\%$ observed on GCP M in Serget and on GCP C in Seth. On GCP E, the 5-6$\%$ water ice rich bright area patch analyzed by \cite{Filacchione2016}, we are observing a 40$\%$ increase of the SSA between August and November 2014. The general increase of the SSA on the Imhotep GCPs is probably a precursor of the morphological changes occurred later in this area where roundish expanding depressions were forming and where exposed water ice appeared on the surface \citep{Groussin2015}.

\begin{table}[h!]
\begin{scriptsize}
\centering  
\begin{tabular}{|c|c|c|c|c@{-}c@{-}c|c@{-}c@{-}c|c@{-}c@{-}c|}      
\hline \hline
GCP & Region & Lon & Lat & \multicolumn{3}{c|} {SSA(0.55 $\mu m$)} & \multicolumn{3}{c|} {$S_{0.5-0.8\mu m}$} & \multicolumn{3}{c|} {$S_{1.0-2.5\mu m}$} \\
 & Name & (deg) & (deg) & MTP6 & MTP7 & MTP9 & MTP6 & MTP7 & MTP9 & MTP6 & MTP7 & MTP9 \\
\hline \hline
A & Ma'at & 16.5 & 24.5 & 0.051 & 0.065 & 0.072 & 1.799 & 1.840 & 1.935 & 0.430 & 0.425 & 0.395 \\							
B & Aker & 56.0 & -12.5 & 0.063 & 0.092 & 0.112 & 2.025 & 1.976 & 2.042 & 0.527 & 0.508 & 0.498 \\							
C & Babi & 80.0 & 58.0 & 0.055 & 0.070 & 0.103 & 1.765 & 1.716 & 1.725 & 0.418 & 0.395 & 0.405  \\							
D & Imhotep & 115.0 & -11.0 & 0.053 & 0.054 & 0.057 & 2.004 & 1.877 & 1.812 & 0.466 & 0.463 & 0.467  \\						
E & Khepry-Ash & 116.0 & 10.5 & 0.037 & 0.047 & 0.052 & 1.859 & 1.614 & 1.594 & 0.419 & 0.396 & 0.330 \\					
F & Ash & 127.0 & 58.0 & 0.050 & 0.053 & 0.060 & 1.859 & 1.840 & 1.845 & 0.467 & 0.408 & 0.499 \\							
G & Imhotep & 133.0 & 5.0 & 0.056 & 0.061 & 0.064 & 1.977 & 1.885 & 1.829 & 0.461 &  0.463 & 0.466 \\     					
H & Ash & 139.0 & 50.0 & 0.051 & 0.055 & 0.057 & 1.937 & 2.209 & 1.922 & 0.468 & 0.391 & 0.474 \\    						
I & Imhotep & 140.0 & -10.0 & 0.061 & 0.063 & 0.068 & 1.971 & 1.913 & 1.704 & 0.477 & 0.504 & 0.460 \\   					
J & Imhotep & 149.5 & -2.0 & 0.064 & 0.090 & 0.097 & 1.976 & 1.891 & 1.457 & 0.454 & 0.459 & 0.461 \\						
K & Seth & 219.5 & 18.0 & 0.051 & 0.064 & 0.082 & 1.920 & 1.926 & 1.890 & 0.464 & 0.448 & 0.445 \\					
L & Anuket & 285.5 & -5.5 & 0.050 & 0.054 & 0.061 & 2.001 & 2.054 & 2.082 & 0.461 & 0.444 & 0.443 \\						
M & Serget & 320.5 & 21.0 & 0.054 & 0.059 & 0.101 & 1.931 & 2.238 & 2.410 & 0.482 & 0.603 & 0.818 \\						
N & Ma'at & 347.0 & 35.0 & 0.055 & 0.059 & 0.070 & 1.769 & 1.817 & 1.990 & 0.435 & 0.436 & 0.449 \\							
\hline \hline
\end{tabular}
\caption{Temporal variation of the SSA(0.55 $\mu m$), visible spectral slope $S_{0.5-0.8 \mu m}$ (in $10^{-3}$ $nm^{-1}$) and infrared spectral slope $S_{1.0-2.5 \mu m}$(in $10^{-3}$ $nm^{-1}$) during mission periods MTP6 (August 2014, heliocentric distance 3.62 AU), MTP7 (September 2014, heliocentric distance 3.31 AU) and MTP9 (November 2014, heliocentric distance 2.93 AU).}  
\label{table:3}
\end{scriptsize}                   
\end{table}

\begin{table}[h!]
\begin{scriptsize}
\centering  
\begin{tabular}{|c|c|c|c|c@{-}c@{-}c|c@{-}c@{-}c|c@{-}c@{-}c|}      
\hline \hline
GCP & Region & Lon & Lat & \multicolumn{3}{c|} {BD(3.2 $\mu m$)} & \multicolumn{3}{c|} {BC(3.2 $\mu m$)} & \multicolumn{3}{c|} {BD(2.0 $\mu m$)} \\
& Name & (deg) & (deg) & MTP6 & MTP7 & MTP9 & MTP6 & MTP7 & MTP9 & MTP6 & MTP7 & MTP9 \\
\hline \hline

A & Ma'at &        16.5 & 24.5 &     12.4 &   10.9 &    12.4 &    3.237 &      3.246 &      3.275 &   0.6 &   0.5 &   1.3 \\
B & Aker &         56.0 & -12.5 &    6.9 &    8.4 &     7.1 &     3.227 &      3.265 &      3.227 &   1.0 &   0.5 &   0.5 \\
C & Babi &         80.0 & 58.0 &     14.8 &   14.4 &    14.8 &    3.237 &      3.246 &      3.246 &   0.6 &   0.5 &   1.1 \\
D & Imhotep &      115.0 & -11.0 &   10.3 &   13.0 &    12.9 &    3.246 &      3.237 &      3.237 &   0.5 &   0.7 &   0.7 \\
E & Khepry-Ash     & 116.0 & 10.5 &    16.2 &   16.3 &    19.7 &    3.246 &      3.246 &      3.246 &   1.3 &   1.2 &   1.6 \\  
F & Ash &         127.0 & 58.0 &    12.8 &   12.3 &    12.5 &    3.237 &      3.237 &      3.227 &   0.6 &   0.1 &   0.6 \\
G & Imhotep &      133.0 & 5.0 &     9.3 &    11.6 &    11.9 &    3.237 &      3.227 &      3.237 &   0.6 &   0.7 &   0.6 \\
H & Ash &          139.0 & 50.0 &    10.4 &   13.1 &    9.3 &     3.237 &      3.256 &      3.218 &   0.5 &   0.1 &   0.5 \\
I & Imhotep &      140.0 & -10.0 &   9.4 &    9.9 &     10.6 &    3.246 &      3.246 &      3.246 &   0.5 &   0.8 &   0.8 \\
J & Imhotep &      149.5 & -2.0 &    9.3 &    10.9 &    10.6 &    3.246 &      3.227 &      3.227 &   0.8 &   0.6 &   0.8 \\
K & Seth        &  219.5 & 18.0 &    12.3 &   11.3 &    10.5 &    3.237 &      3.237 &      3.246 &   0.5 &   0.3 &   0.3 \\
L & Anuket &       285.5 & -5.5 &    13.3 &   12.6 &    15.5 &    3.237 &      3.237 &      3.218 &   0.5 &   0.6 &   0.3 \\
M & Serget &       320.5 & 21.0 &    9.4 &    6.9 &     6.9 &     3.246 &      3.444 &      3.341 &   0.5 &   0.2 &   0.2 \\
N & Ma'at &        347.0 & 35.0 &    12.5 &   14.7 &    13.2 &    3.237 &      3.227 &      3.237 &   0.6 &   0.4 &   0.7 \\
\hline \hline
\end{tabular}
\caption{Temporal variation of the 3.2 $\mu m$ organic material band depth (in $\%$), band center (in $\mu m$) and water ice 2.0 $\mu m$ band depth (in $\%$) during mission periods MTP6 (August 2014, heliocentric distance 3.62 AU), MTP7 (September 2014, heliocentric distance 3.31 AU) and MTP9 (November 2014, heliocentric distance 2.93 AU).}  
\label{table:4}
\end{scriptsize}                   
\end{table}

\subsection{Visible spectral slope $S_{0.5-0.8 \mu m}$ cylindrical map}

As mentioned before, 67P/CG nucleus reflectance spectra show an intense reddening in the 0.5-0.8 $\mu m$ spectral range. In both VIRTIS-M and OSIRIS data, it has been verified that the reddening is phase-dependent, with a significant increase at high solar phases \citep{Ciarniello2015, Fornasier2015}. 

\begin{figure}[h!]
	\centering
		\includegraphics[width=17cm]{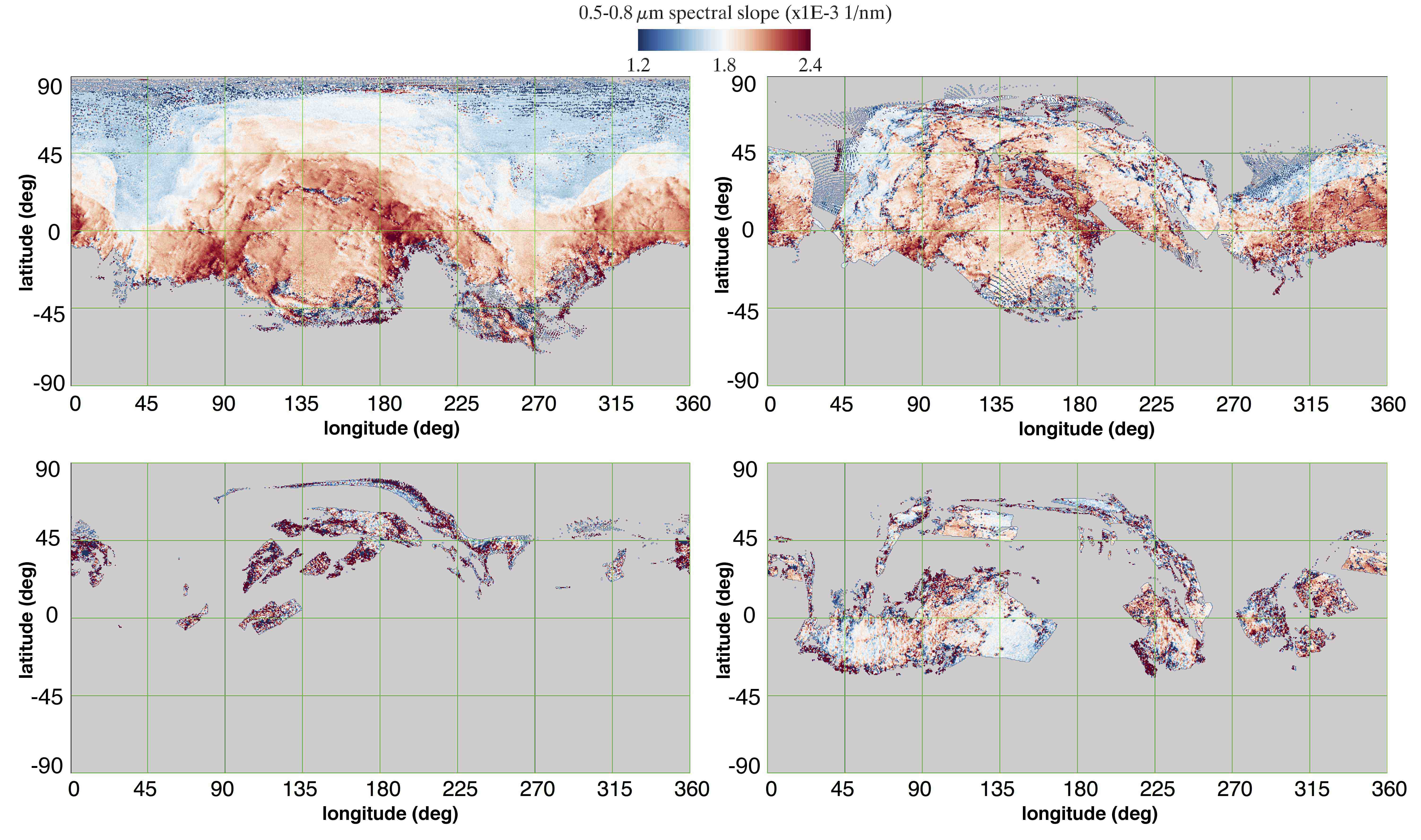}
		\caption{Visible spectral slope $S_{0.5-0.8 \mu m}$ cylindrical maps for MTP6 (top left), MTP7 (top right), MTP8 (bottom left), MTP9 (bottom right).} 
		\label{fig:figure11}
\end{figure}

Within the limits of the photometric correction \citep{Ciarniello2015} and geometric reconstruction applied to derive reflectance spectra, the spectral slope $S_{0.5-0.8\mu m}$ maps shown in Fig. \ref{fig:figure11} are corrected for illumination/viewing geometry and local topography. The resulting distribution of the $S_{0.5-0.8 \mu m}$ shows a variability between 1.2 $\cdot 10^{-3}$ and 2.4 $\cdot 10^{-3} \ nm^{-1}$. On the MTP6 maps we observe that the smallest values, between 1.2 $\cdot 10^{-3}$ and 1.7 $\cdot 10^{-3} \ nm^{-1}$ are measured above the active Hapi region and are rendered with the blue color. Intermediate values of about 1.8 $\cdot 10^{-3} \ nm^{-1}$ are observed in the Ma'at and Hathor regions on the small lobe and across Seth, Ash and Babi on the body. In general the slope is greater than 1.9 $\cdot 10^{-3} \ nm^{-1}$ on the remaining regions, reaching the maximum value of 2.4 $\cdot 10^{-3} \ nm^{-1}$ in the Hatmehit region of the small lobe and on two places on the body, Khepry and on the Apis-Atum-Ash boundary, rendered in the red color. Some points in correspondence of the basis of some elevated structures placed around the Imhotep plain show a decrease of the slope, which changes from the average 1.9 $\cdot 10^{-3} \ nm^{-1}$ measured on the central plain to less than 1.3 $\cdot 10^{-3} \ nm^{-1}$. While for some of these points this effect is probably caused by shadows due to local topography, on others the drop seems to be correlated with water ice-rich patches, like on GCP E. 
Similarly to the trend observed for the SSA, moving towards perihelion from MTP6 to MTP9 (corresponding to a time-interval of about 4 months or 0.69 AU in heliocentric distance), the spectral slope shows local changes: from the comparison of the maps in Fig. \ref{fig:figure11} and from the analysis of the 14 GCPs reported in Table \ref{table:3}, it turns out that on some regions, like Seth and Imhotep the reddening has a maximum decrease of about 25$\%$. On other regions, like Ma'at and Serget, a significant increase, up to 7$\%$ and $25\%$, respectively, is observed. Finally, the slope remains almost constant on Aker and Ash.  
\cite{Ciarniello2015} and \cite{Fornasier2015} have also made a comparison of the VIS spectrophotometric parameters measured  by VIRTIS and OSIRIS, respectively, on 67P/CG and previously observed comets, like 1P/Halley, 19P/Borrelly, 103P/Wild, 9P/Tempel and 81P/Wild. A direct comparison among these objects is quite difficult because different space missions have collected data at different heliocentric distances and with variable phase angle coverages. Certainly Rosetta data clearly show that 67P/CG spectral slopes vary across the surface and are dynamical quantities changing with time depending upon the heliocentric distance and level of the gaseous activity.  

\subsection{Infrared spectral slope $S_{1.0-2.5 \mu m}$ cylindrical map}

\begin{figure}[h!]
	\centering
		\includegraphics[width=17cm]{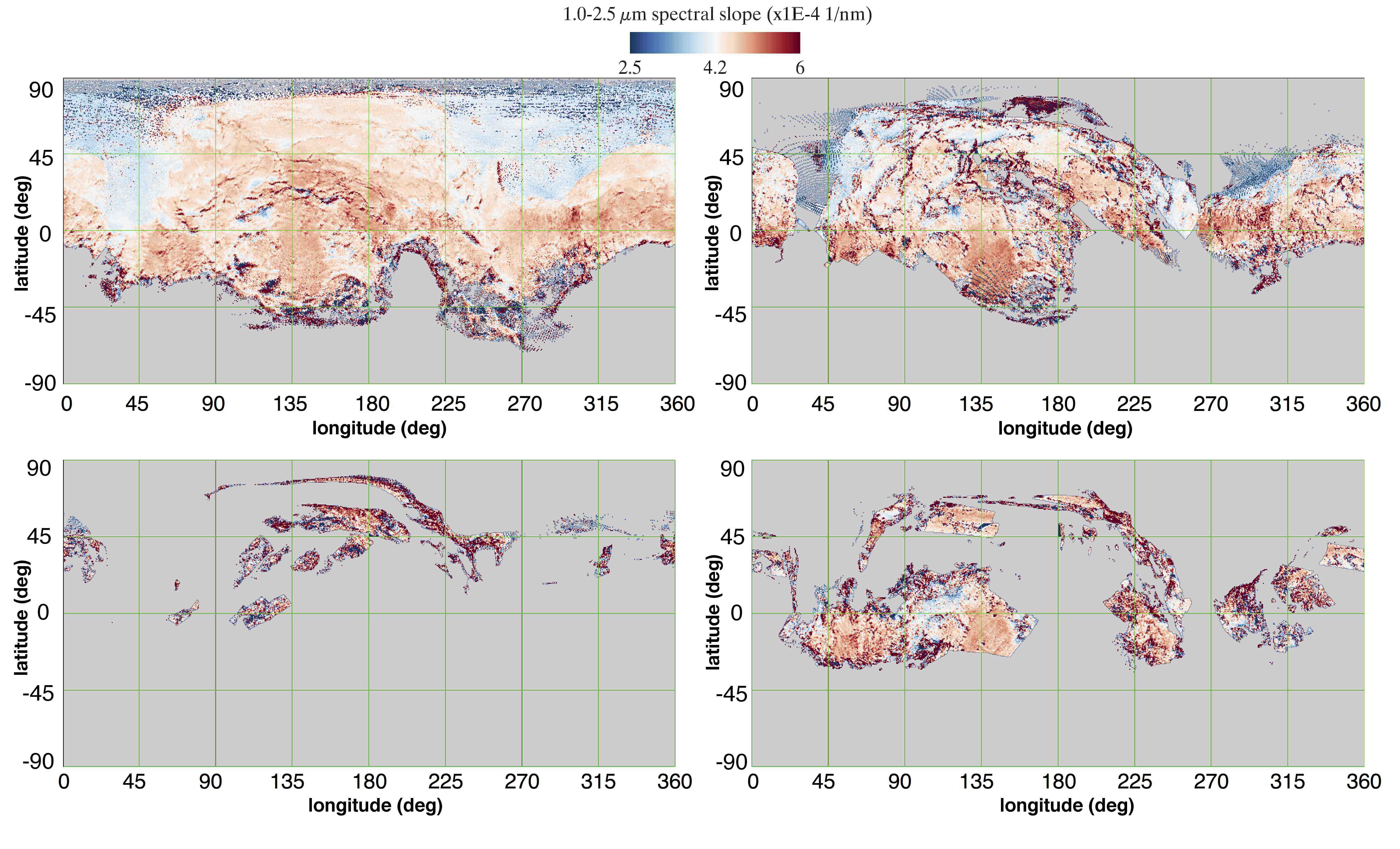}
		\caption{Infrared spectral slope $S_{1.0-2.5 \mu m}$ cylindrical maps for MTP6 (top left), MTP7 (top right), MTP8 (bottom left), MTP9 (bottom right).} 
		\label{fig:figure12}
\end{figure}

The measured infrared slope $S_{1.0-2.5 \mu m}$ varies between 2.5 and 6.0 $\cdot 10^{-4} \  nm^{-1}$. The distribution of the slope on the surface as shown in Fig. \ref{fig:figure12} resembles the visible slope one: lower values are observed on Hapi and around the Imhotep plain, in particular around GCP E area. The strongest reddening, equal to 6 $\cdot 10^{-4} \ nm^{-1}$, is measured in MTP6 on Anuket region, towards the Hathor and Serget boundary.
Moving from MTP6 to MTP9 the $S_{1.0-2.5 \mu m}$ slope values remain almost constant (see Table \ref{table:3}) with the exclusion of Serget and GCP E, where an increase of $70\%$ and a decrease of about $21\%$ have been observed, respectively. While the effect on Serget could be caused by photometric correction residuals, the blueing observed on the western side of the Imhotep plain towards GCP E seems to be caused by the exposure at the surface of fresh material. On the same region we observe a similar decrease of the $S_{0.5-0.8 \mu m}$ slope together with an increase of the SSA(0.55 $\mu m$). A further verification of this effect occurring on GCP E will be given later in the organic material and water ice bands spectral indicators discussion.
 
\subsection{3.2 $\mu m$ organic material absorption band cylindrical map}
\begin{figure}[h!]
	\centering
		\includegraphics[width=17cm]{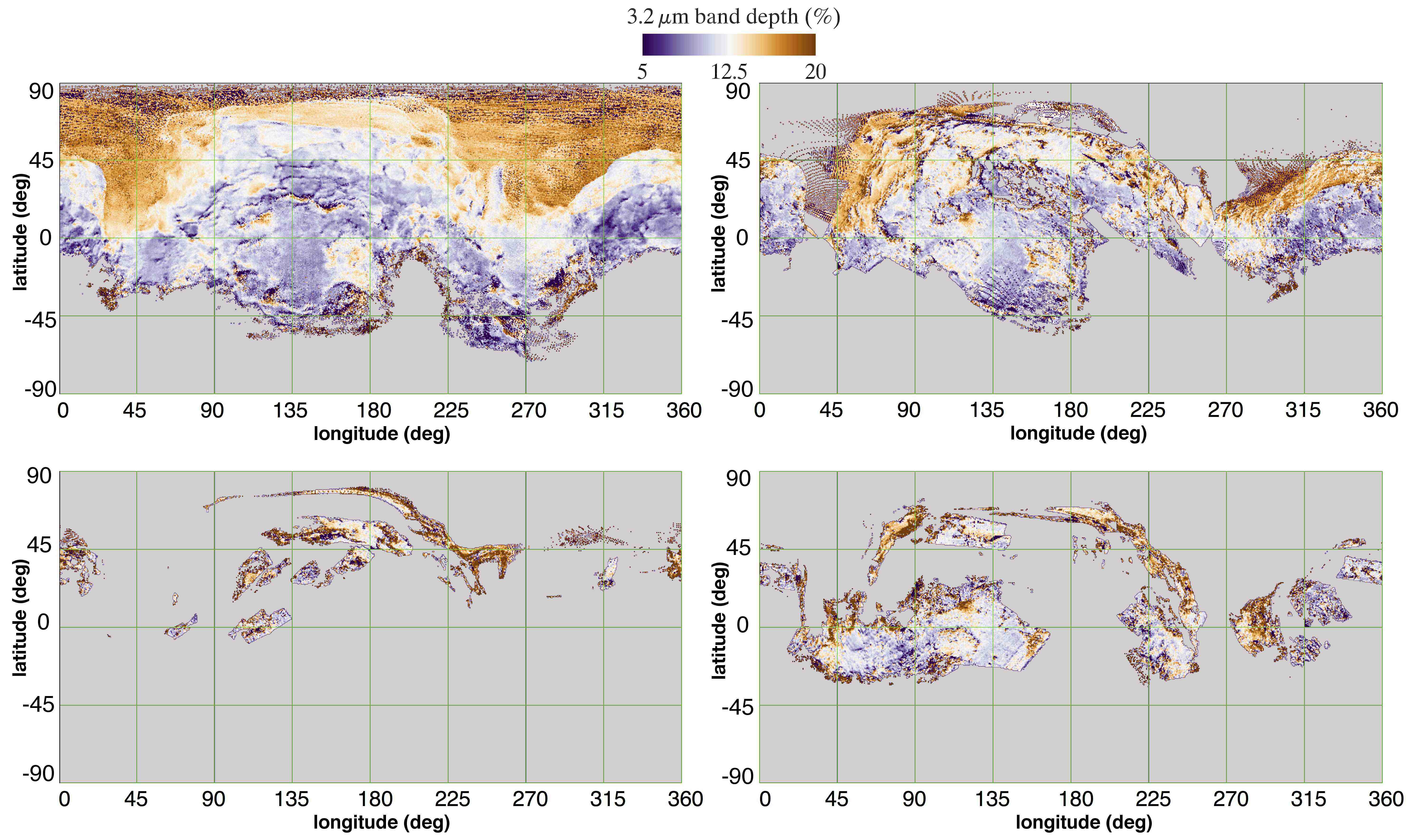}
		\caption{Organic material 3.2 $\mu m$ band depth cylindrical maps for MTP6 (top left), MTP7 (top right), MTP8 (bottom left), MTP9 (bottom right).} 
		\label{fig:figure13}
\end{figure}

\begin{figure}[h!]
	\centering
		\includegraphics[width=17cm]{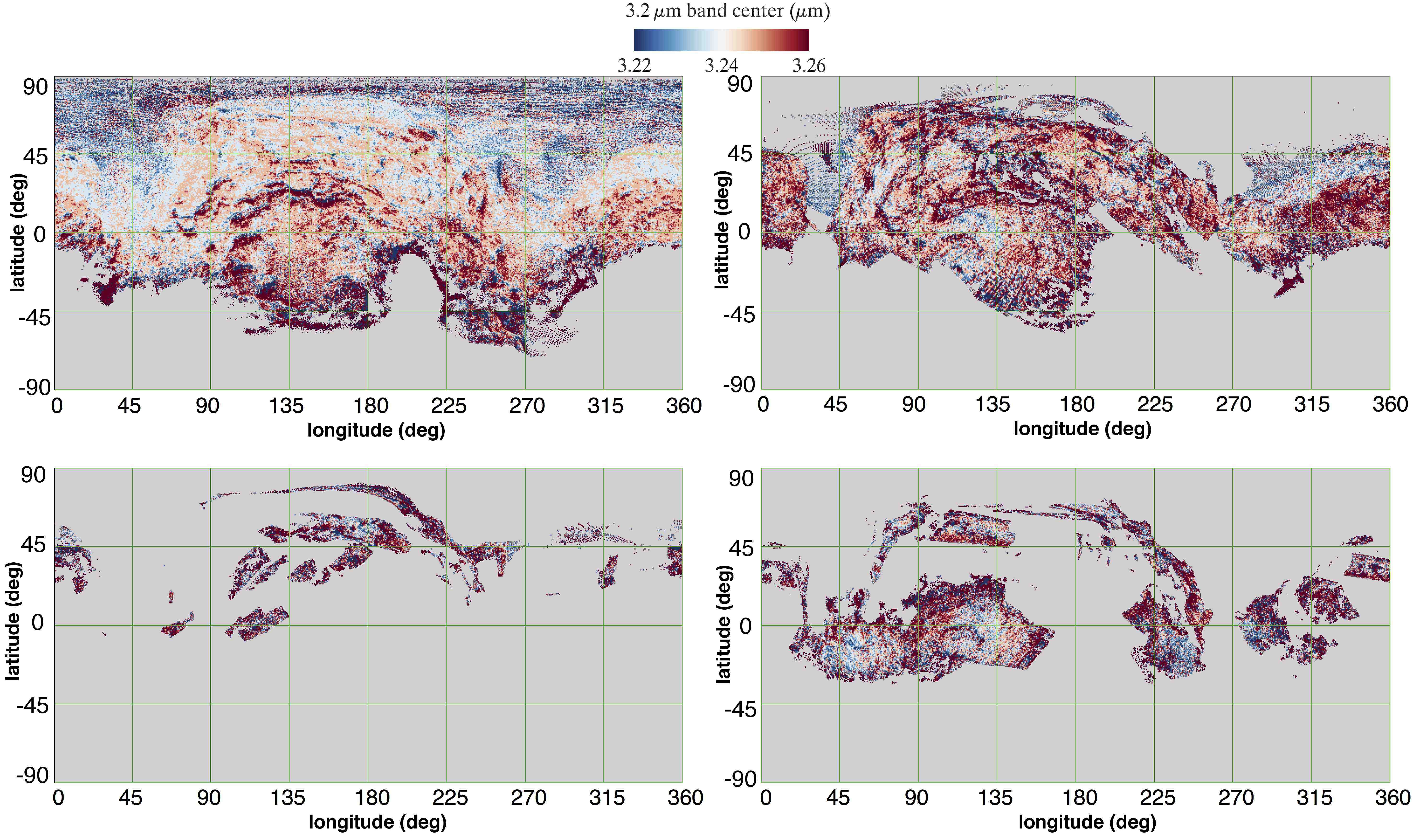}
		\caption{Organic material 3.2 $\mu m$ band center cylindrical maps for MTP6 (top left), MTP7 (top right), MTP8 (bottom left), MTP9 (bottom right).} 
		\label{fig:figure14}
\end{figure}

The  3.2 $\mu m$ band used to trace organics is analyzed and mapped by using both band depth (Fig. \ref{fig:figure13}) and center (Fig. \ref{fig:figure14}). The 3.2 $\mu m$ band depth varies between 5 and 20 $\%$: the highest values are observed above Hapi active region and in several patches in Seth, Ash and Imhotep. The minimum value is measured on the small lobe around the Hatmehit depression and in Nut, Ma'at, Serget, Maftet and Bastet regions. In general the correlation between organic material band depth and spectral slopes is well-evident: regions with higher band depth show the lowest visible and infrared spectral slopes. This property is further discussed later in Fig. \ref{fig:figure18}. Since we have several evidences indicating the presence of water ice in the Hapi region, we have verified how it could affect the 3.2 $\mu m$ band depth increase measured on this region. The presence of the H-O-H vibrations (symmetric and antisymmetric modes) at 2.7-2.8 $\mu m$ in fact partially overlaps the organic signature, changing its shape and depth.

In order to disentangle the water from organic material abundance, the 3.2 $\mu m$ band center has been computed and mapped in Fig. \ref{fig:figure14}. The band center position shows a variability mostly in the 3.22-3.44 $\mu m$ range: the shortest wavelength position is observed mainly on Hapi and Anuket in the active units of the neck/north pole while it shifts to longer wavelengths on a great part of the small lobe and on the meridional hemisphere. This is a clear indication that the presence of water ice causes an increase of the 3.2 $\mu m$ band depth and a shift of the band center towards shorter wavelengths. A similar behavior is observed in fact not only in Hapi but also on the deposits seen around Imothep, like GCP E and on the hills of the eastern side of Imhotep plain, the same places were \cite{Filacchione2016} have observed surface water ice.  
The temporal trends of the 3.2 $\mu m$ band depth and center above the fourteen GCPs reported in Table \ref{table:4} show that the band depth has increased from MTP6 to MTP9 on Imhotep (GCPs D, G, I, J), on Khepry-Ash (E), Anuket (L) and Ma'at (N).
Conversely, the band depth has decreased on Serget (GCP M) and Ash (H). The band center appears in general more stable.  

\subsection{2.0 $\mu m$ water ice band cylindrical map}
\begin{figure}[h!]
	\centering
		\includegraphics[width=17cm]{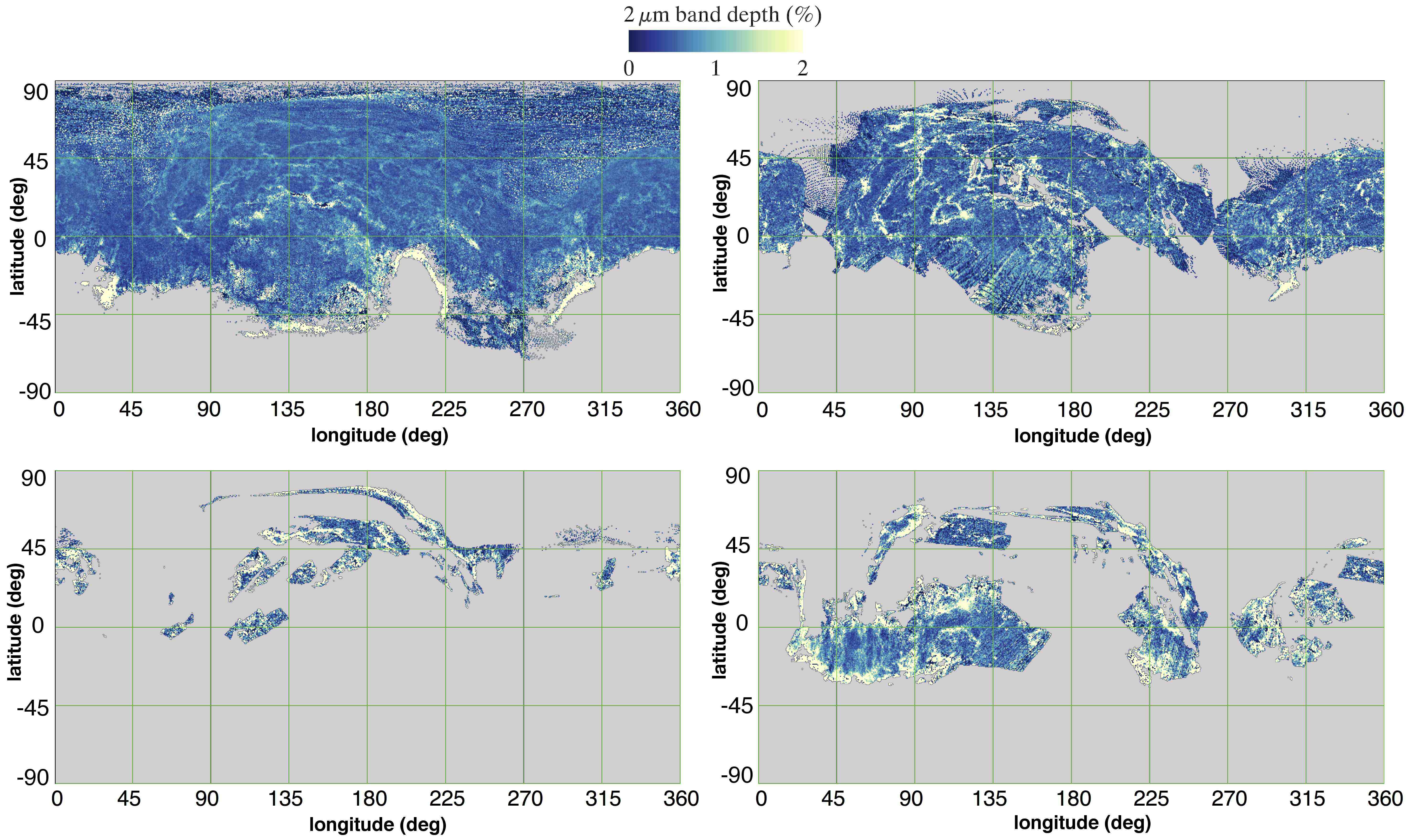}
		\caption{Water ice 2.0 $\mu m$ band center cylindrical maps for MTP6 (top left), MTP7 (top right), MTP8 (bottom left), MTP9 (bottom right).} 
		\label{fig:figure15}
\end{figure}

Very small amounts of water ice are in general visible on the surface of 67P/CG. When present, the ice is in form of transient condensed frost following the diurnal cycle \citep{DeSanctis2015} or is in intimate and areal mixing within the ubiquitous dark terrain, like on the bright albedo patches seen in Imhotep \citep{Filacchione2016}. A more systematic search for water ice deposits has been performed by measuring and mapping the depth of the water ice 2.0 $\mu m$ band (Fig. \ref{fig:figure15}). Apart from some points located along the south hemisphere terminator, where the band depth retrieval is affected by the local rough topography and by extreme angles of illumination, we observe a small band depth variability, between 0 and 2$\%$. The 2$\%$ maximum value is detected around Imhotep central plain, in particular on GCP E. On the eastern region of Imhotep a wide area located at about 170$^\circ$$\le$lon$\le$180$^\circ$, -30$^\circ$$\le$lat$\le$10$^\circ$ shows a band depth $\ge 1 \%$. This is the same area where the peculiar roundish features have been identified by \cite{Auger2015}. The GCP E place shows an evident band depth increase from MTP6 to MTP9 (Table \ref{table:4}). We observe a general increase of the abundance of water ice in specific areas going from MTP6 to MTP9.

Given the shallow depth of the absorption feature, the differences observed in the water ice band depth distribution during the different MTPs could be caused by changes in illumination and viewing geometries: the MTP6 map shows very small contrast, due to the fact that it was acquired at constant phase angle (30$^\circ$-40$^\circ$). The MTP7 map appears more contrasted around the elevated structures, as for example in Seth and Ash, because these observations were taken at local solar times corresponding to mid morning and mid afternoon, resulting in shadows casted on opposite sides of the local elevated structures. The illumination is even more extreme in MTP9, when the spacecraft was orbiting on a terminator orbit at 90$^\circ$ phase angle.

Although the photometric correction removes any variability caused by changing illumination conditions, the role of topography and consequent local long shadows becomes more relevant when observations are taken at high incidence angles. This is the case of data acquired at phase angles up to 90$^\circ$ in MTP7, 8, 9 as compared to MTP6 which was acquired at approximately constant phase angles  of 30-40$^\circ$. This residual effect is indeed observed in Fig. \ref{fig:figure15} (and somewhat also in Fig. \ref{fig:figure12}), probably due to the limited range of band depth, never exceeding the 2$\%$ level, we are dealing with.

\section{CSI correlation with geomorphological classes}
\begin{figure}[h!]
	\centering
		\includegraphics[width=17cm]{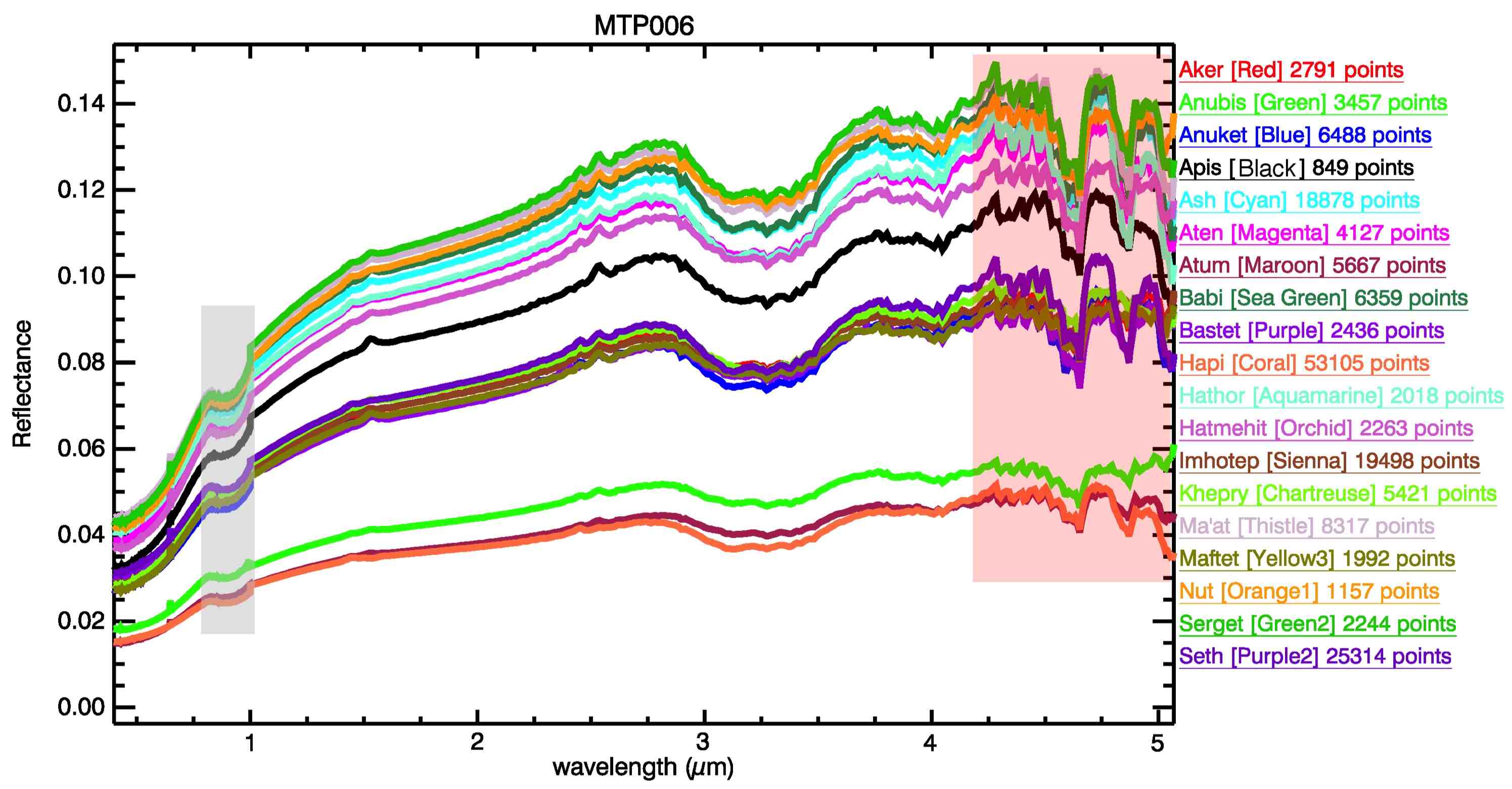}
		\caption{Average VIS-IR reflectance spectra of the nineteen morphological regions. For each region are indicated the number of points, corresponding to $0.5 \times 0.5^{\circ}$ bins on cylindrical map projection, used to compute the average spectra. In gray the visible spectral range affected by low signal to noise conditions. In red the infrared range affected by thermal emission removal method, calibration residuals and detector's non-linearity response.} 
		\label{fig:figure16}
\end{figure}

A total of nineteen geomorphological regions have been identified by means of anaglyphs derived from OSIRIS images taken from close orbits \citep{Thomas2015, El-Maarry2015}. These regions show distinctive properties which can be summarized in three major classes: 1) consolidated regions (CR), including strongly consolidated (SC) and brittle regions (BR); 2) non-consolidated regions (NCR), including smooth terrains (ST) and dust-covered brittle (DCB) regions; and 3) large irregular depressions (D).
A great part of the nucleus appears consolidated and fractured. On both comet lobes dust-covered terrains are visible, possibly pointing to the presence of dust transport mechanisms caused by activity. Brittle regions are associated with local mechanical stresses caused by the rotational state of the irregular-shaped nucleus. Conversely, two large irregular depressions appear dust-clean and show a morphology compatible with explosive and mass ejection processes, due to violent sublimation and outgassing. The distribution of the morphological regions across the 67P/CG surface is shown in Fig. \ref{fig:figure10}.
\
For each of these regions, average VIRTIS-M spectra were derived with the aim to explore possible connections between surface morphology and spectral properties. The analysis is performed on MTP6 period data which has the most extensive spatial coverage. The average spectral reflectances in each of the morphological regions are shown in Fig. \ref{fig:figure16}. The features shown in the 4.2-5.0 $\mu m$ spectral range are still uncertain, these wavelengths being affected by the thermal emission removal method, calibration residuals and the detector's non-linear response  as discussed in section 3. For each region a set of spectral indicators, consisting in SSA(0.55 $\mu m$), S$_{0.5-0.8 \ \mu m}$, S$_{1.0-2.5 \ \mu m}$, BC(3.2 $\mu m$) and BD(3.2 $\mu m$) are computed and the resulting values are listed in Table \ref{table:5}.   

\begin{table}
\begin{scriptsize}
\centering  
\begin{tabular}{|c|c|c|c|c|c|c|c|c|c|}      
\hline\hline
Region & SSA(0.55 $\mu m$) &  S$_{0.5-0.8 \ \mu m}$  & S$_{1.0-2.5 \ \mu m}$  & BC(3.2 $\mu m$) & BD(3.2 $\mu m$) & Terrain  & Class & Nr.\\
Name &  &  ($10^{-3}$ $nm^{-1}$) & ($10^{-3}$ $nm^{-1}$) & ($\mu m$) & ($\%$) & Type & ID & Bins \\
\hline\hline
Aker   & 0.050$\pm$0.012 &  1.883$\pm$0.223 & 0.473$\pm$0.068 & 3.191$\pm$0.134 & 0.097$\pm$0.037 & SC & Green & 2791 \\
Anubis & 0.043$\pm$0.017 &  1.684$\pm$0.330 & 0.406$\pm$0.084 & 2.814$\pm$0.147 & 0.084$\pm$0.030 & NCS & Green & 3457 \\
Anuket & 0.045$\pm$0.014 &  1.856$\pm$0.252 & 0.448$\pm$0.107 & 3.075$\pm$0.123 & 0.116$\pm$0.042 & SC & Magenta & 6488 \\
Apis   & 0.050$\pm$0.016 &  2.116$\pm$0.240 & 0.444$\pm$0.070 & 3.113$\pm$0.144 & 0.102$\pm$0.040 & SC & Magenta & 849 \\
Ash    & 0.049$\pm$0.021 &  1.935$\pm$0.295 & 0.464$\pm$0.139 & 3.234$\pm$0.150 & 0.110$\pm$0.041 & NCD & Magenta & 18878 \\
Aten   & 0.051$\pm$0.016 &  1.965$\pm$0.332 & 0.458$\pm$0.103 & 3.243$\pm$0.146 & 0.114$\pm$0.039 & D & Magenta & 4127 \\
Atum   & 0.038$\pm$0.017 &  1.560$\pm$0.314 & 0.365$\pm$0.114 & 2.538$\pm$0.115 & 0.084$\pm$0.032 & SC & Green & 5667 \\
Babi   & 0.053$\pm$0.019 &  1.850$\pm$0.255 & 0.451$\pm$0.115 & 3.239$\pm$0.148 & 0.123$\pm$0.042 & DCB & Blue & 6359 \\
Bastet & 0.048$\pm$0.020 &  1.834$\pm$0.319 & 0.428$\pm$0.152 & 3.056$\pm$0.186 & 0.092$\pm$0.050 & SC & Green & 2436 \\
Hapi   & 0.048$\pm$0.012 &  1.360$\pm$0.179 & 0.339$\pm$0.070 & 2.664$\pm$0.079 & 0.130$\pm$0.029  & NCS & Blue & 53105 \\
Hathor & 0.051$\pm$0.020 &  1.793$\pm$0.358 & 0.441$\pm$0.181 & 3.241$\pm$0.164 & 0.131$\pm$0.052 & SC & Blue & 2018 \\
Hatmehit & 0.052$\pm$0.020 & 2.068$\pm$0.196 & 0.475$\pm$0.060 & 3.210$\pm$0.153 & 0.095$\pm$0.031 & D & Green & 2263 \\
Imhotep & 0.053$\pm$0.014 &  1.955$\pm$0.252 & 0.460$\pm$0.076 & 3.153$\pm$0.111 & 0.103$\pm$0.038 & NCS & Magenta & 19498 \\
Khepry  & 0.052$\pm$0.016 &  2.004$\pm$0.272 & 0.441$\pm$0.077 & 3.076$\pm$0.132 & 0.104$\pm$0.039 & SC & Magenta & 5421 \\
Ma'at   & 0.052$\pm$0.017 &  1.843$\pm$0.283 & 0.448$\pm$0.107 & 3.240$\pm$0.154 & 0.117$\pm$0.044 & NCD & Magenta & 8317 \\
Maftet  & 0.048$\pm$0.016 &  1.980$\pm$0.283 & 0.452$\pm$0.079 & 3.070$\pm$0.172 & 0.087$\pm$0.048 & SC & Green & 1992 \\
Nut     & 0.054$\pm$0.014 &  1.998$\pm$0.184 & 0.484$\pm$0.059 & 3.248$\pm$0.138 & 0.090$\pm$0.032 & D & Green & 1157 \\
Serget  & 0.053$\pm$0.017 &  1.941$\pm$0.251 & 0.466$\pm$0.081 & 3.242$\pm$0.169 & 0.104$\pm$0.040 & SC & Magenta & 2244 \\
Seth    & 0.052$\pm$0.020 &  1.734$\pm$0.266 & 0.438$\pm$0.010 & 3.201$\pm$0.092 & 0.133$\pm$0.036 & WCB & Blue & 25314 \\
\hline \hline
\end{tabular}
\end{scriptsize}
\caption{Average spectral indicators and standard deviations calculated for morphological classes during the MTP6 period at heliocentric distance of 3.6 AU. Terrain type legend: SC=Strongly Consolidated; NCS=Non Consolidated Smooth; NCD=Non Consolidated Dust; DCB=Dust Covered Brittle; WCB=Weakly consolidated brittle; D=Depression. Morphological classes from \cite{Thomas2015, El-Maarry2015}. Class ID color-code is based on VIRTIS-M classification.}  
\label{table:5}                   
\end{table}

\begin{figure}[h!]
	\centering
		\includegraphics[width=18cm]{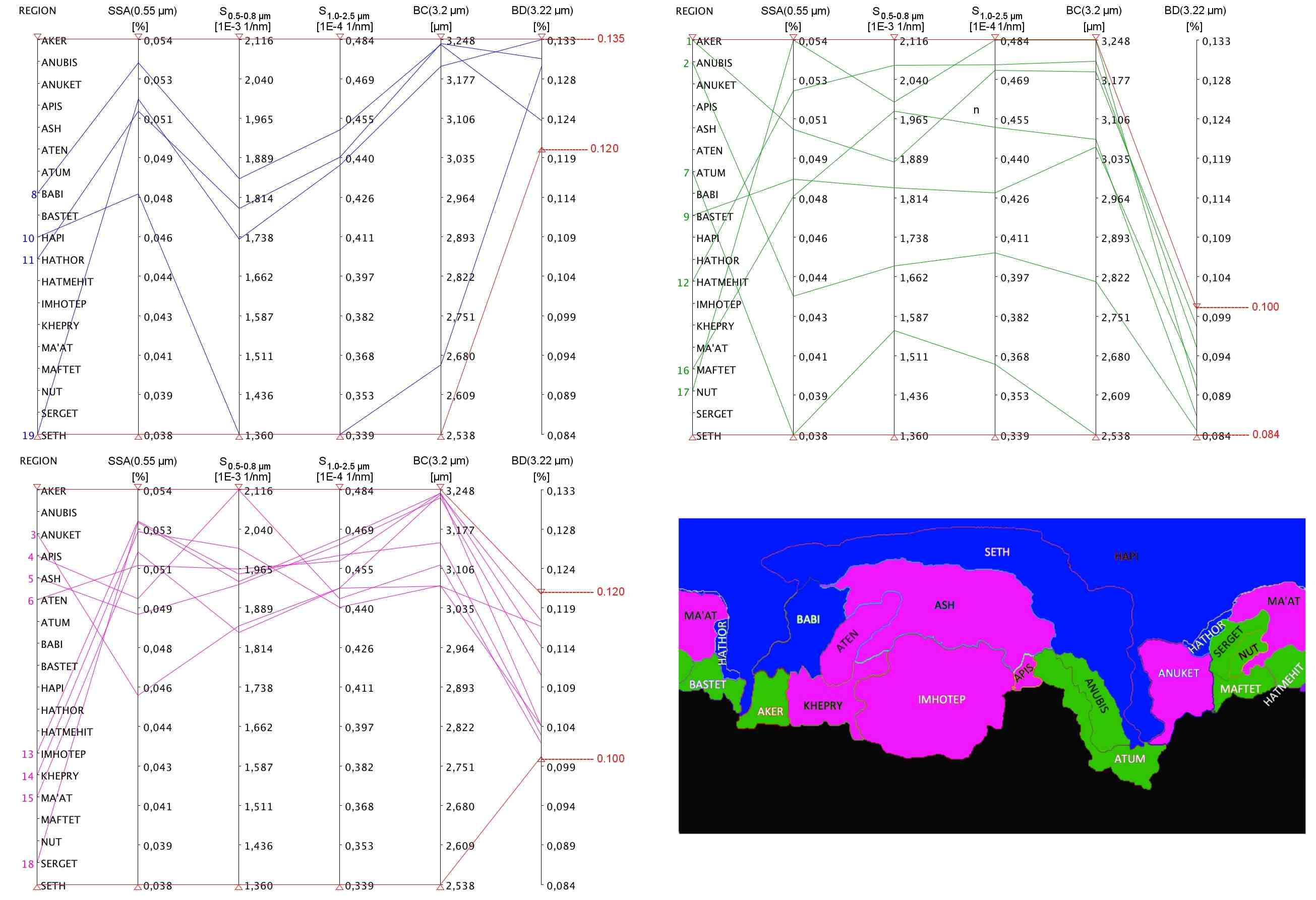}
		\caption{Classification of the nineteen regions' spectral parameters by means of parallel coordinate method. Top left panel: smooth/active regions spectral class (blue points) corresponding to BD(3.2 $\mu m$)$>0.12\%$. Top right panel: depressions regions spectral class (green points) corresponding to BD(3.2 $\mu m$)$<0.10\%$. Bottom left panel: dust covered regions spectral class (magenta points) corresponding to 0.10$\% \le$ BD(3.2 $\mu m$)$\% \le 0.12\%$. Bottom right panel: cylindrical map with the distribution of the three spectral classes as derived by the parallel coordinate method.} 
		\label{fig:figure17}
\end{figure}

In order to explore the properties and define homogeneous classes among the morphological classes, we have analyzed the dataset by means of the Parallel Coordinates method \citep{Inselberg1985}. Parallel Coordinates is a visualization technique widely used to explore multivariate datasets with high-dimensional geometry. The theory of the point-line duality at the base of the method has been generalized for high dimensions allowing to perform classification, clustering and regression on n-dimensional datasets \citep{Inselberg2009}. We have applied a parallel coordinate scheme to explore correlations among geomorphological units average spectral indicators as listed in Table \ref{table:5}. Each spectral indicator is visualized with a vertical axis. The ensemble of the five indicators gives therefore five parallel coordinates axes, plus a sixth axis corresponding to the region name. In this way a six-dimensional space containing region names, SSA(0.55 $\mu m$), S$_{0.5-0.8 \ \mu m}$,  S$_{1.0-2.5 \ \mu m}$, BC(3.2 $\mu m$) and BD(3.2 $\mu m$) has been defined. The numerical values of the spectral indicators corresponding to each class are connected by a polyline which shows the variability of the n-th region in the 6th dimensional space. The resulting representation is shown in Fig. \ref{fig:figure17} where the polylines are color-coded to match the variability of the BD(3.2 $\mu m$) which is chosen to drive the classification: three clusters are defined according to the BD(3.2 $\mu m$) values, corresponding to values $<0.10\%$ (coded with green lines, top right panel), 0.10-0.12$\%$ (magenta lines, bottom left panel), $>0.12\%$ (blue lines, top left panel). We found that these classes quite nicely correspond to depressions, dust covered areas and smooth/active areas, respectively, according to the \cite{Thomas2015, El-Maarry2015} geomorphological study. Consolidated terrains are characterized by highly variable and uncorrelated spectral parameters and therefore appear distributed across all three classes. The resulting spatial distribution on the surface for the three classes defined on the basis of the BD(3.2 $\mu m$) average values is shown in Fig. \ref{fig:figure17}, bottom right panel. 

\begin{figure}[h!]
	\centering
		\includegraphics[width=18cm]{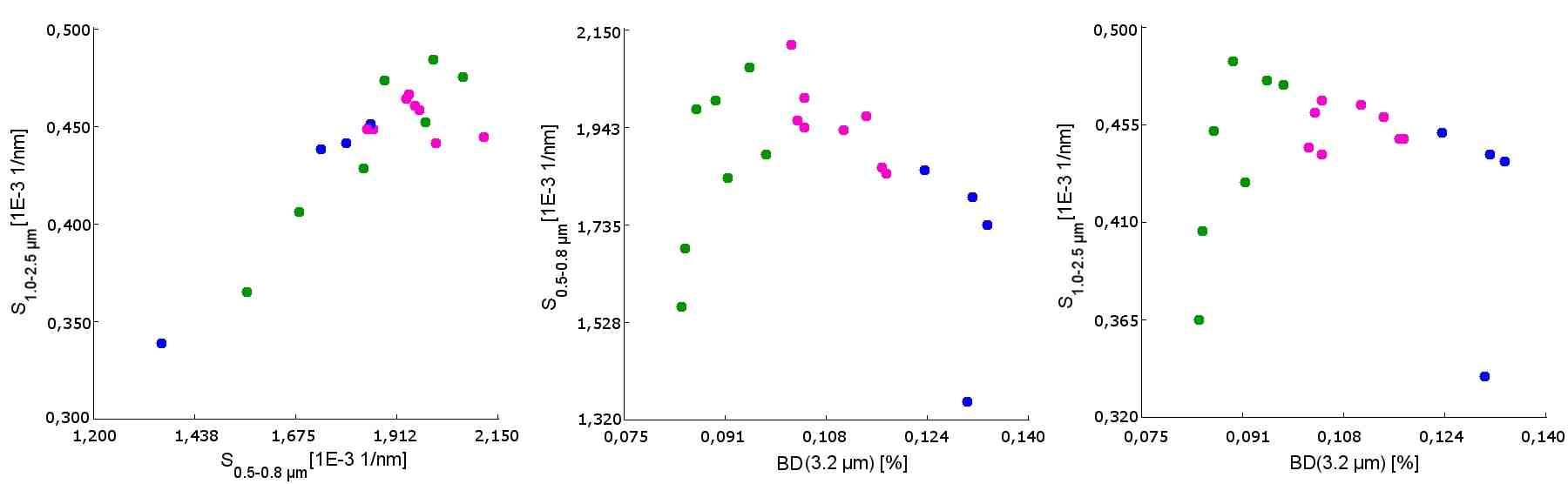}
		\caption{Scatterplots of the three spectral classes as derived by the parallel coordinate method. Left panel: scatterplot S$_{0.5-0.8 \ \mu m}$ vs. S$_{1.0-2.5 \ \mu m}$. Central panel: scatterplot BD(3.2$ \mu m$) vs. S$_{0.5-0.8 \ \mu m}$. Right panel: scatterplot BD(3.2$ \mu m$) vs. S$_{1.0-2.5 \ \mu m}$.} 
		\label{fig:figure18}
\end{figure}

In general we observe a strong correlation between the S$_{0.5-0.8 \ \mu m}$ and S$_{1.0-2.5 \ \mu m}$ values, whose scatterplot shows that reddening increases in both visible and infrared ranges at the same time (Fig. \ref{fig:figure18}, left panel). However, the distribution in the three classes appears to be not directly connected with slope. In particular the green class encompass the whole range of slope values, and the blue point corresponding to Hapi, close to the origin of the scatterplot, appears disconnected from the rest of the blue class. 
Conversely, the BD(3.2 $\mu m$) vs. S$_{0.5-0.8 \ \mu m}$ scatterplot in Fig. \ref{fig:figure17}, central panel, and the BD(3.2 $\mu m$) vs. S$_{1.0-2.5 \ \mu m}$ scatterplot in Fig. \ref{fig:figure17}, right panel, both show an increase of the slope values with BD(3.2 $\mu m$) within the green class (depressions), a slightly decreasing slope in the magenta class (dust covered areas) and a stronger decrease of the slopes in the blue class (smooth/active areas). VIRTIS-M data clearly indicate that the regions with the highest 3.2 $\mu m$ band depth do not show a similar maximum in S$_{0.5-0.8 \ \mu m}$ slope, as we would expect from organic material rich units. This indeed seems to be the consequence of the presence of the H$_2$O absorption in the blue class active regions of the neck and north pole region. Since H$_2$O absorption overlaps the organic signature, this affects the measured band depth resulting in greater values. 
This effect has been already reported on water ice-rich areas by \cite{Filacchione2016} where the 3.2 $\mu m$ band, ubiquitously observed throughout the cometary nucleus and interpreted \citep{Sunshine2012, Capaccioni2015a} to be caused by the presence of organic compounds, appears deformed towards shorter wavelengths by the strong 3.0 $\mu m$ signature of water ice.
A further indication of the presence of H$_2$O in the blue class regions is given by the 3.2 $\mu m$ band center map in Fig.\ref{fig:figure14} which shows a shift towards shorter wavelengths in these areas.

\begin{figure}[h!]
	\centering
		\includegraphics[width=16cm]{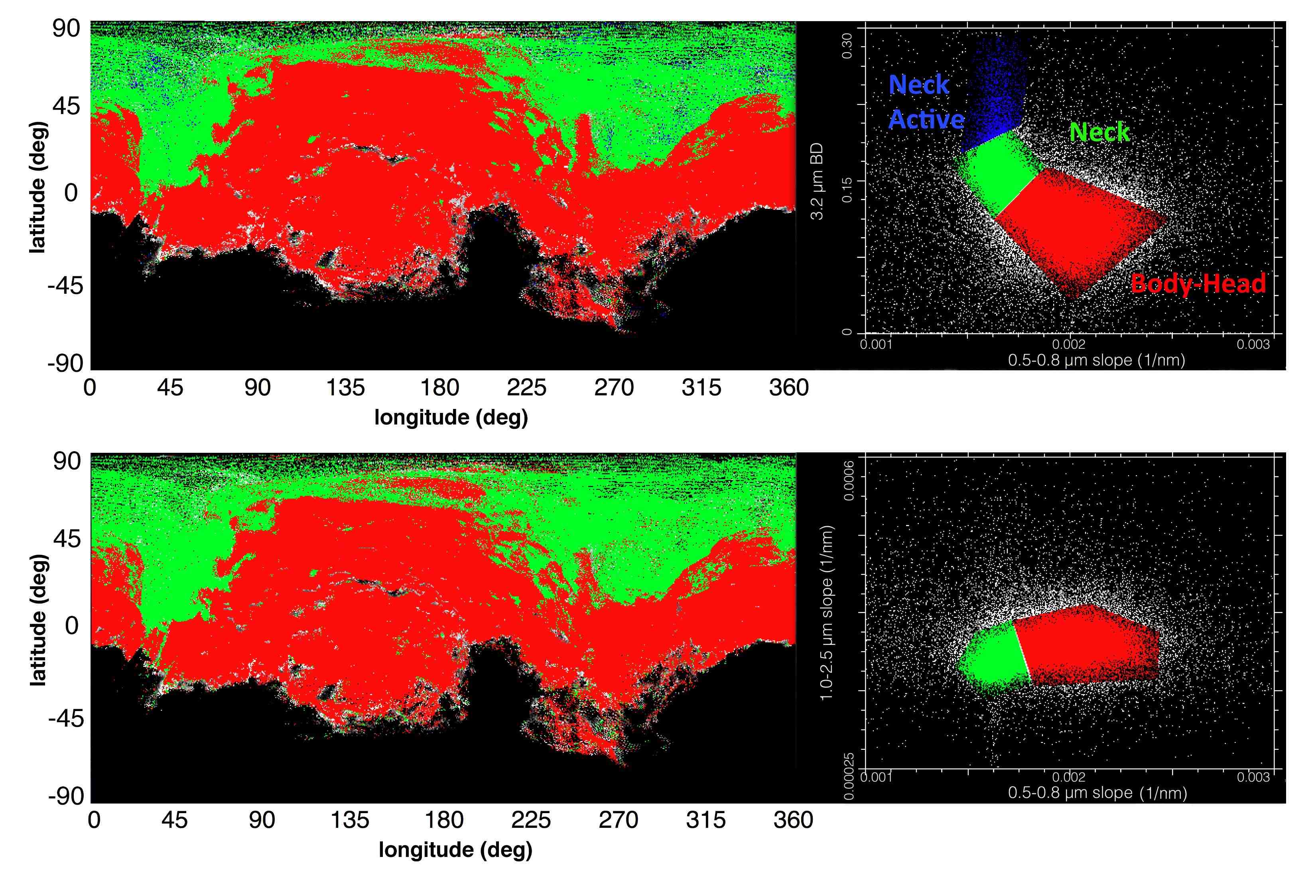}
		\caption{Top panel: S$_{0.5-0.8 \ \mu m}$ vs. BD(3.2 $\mu m$) scatterplot and classes distribution in cylindrical projection. Bottom panel: S$_{0.5-0.8 \ \mu m}$ vs. S$_{1.0-2.5 \ \mu m}$ scatterplot and classes distribution in cylindrical projection. Both panels refer to MTP6 data.} 
		\label{fig:figure19}
\end{figure}

Finally, the distribution of the points in the S$_{0.5-0.8 \ \mu m}$ vs. BD(3.2 $\mu m$) and in the  S$_{0.5-0.8 \ \mu m}$ vs. S$_{1.0-2.5 \ \mu m}$ scatterplots shown in Fig. \ref{fig:figure19} evidences a well-defined separation between active areas in the neck/north pole region (green and blue classes) and the rest of the body (red class). The resulting classification is basically the same by using the two couples of spectral indicators, with the S$_{0.5-0.8 \ \mu m}$ vs. BD(3.2 $\mu m$) scatterplot offering the advantage to identify also active areas in the neck.  

\section{Conclusions}
For the first time VIRTIS-M aboard Rosetta had the opportunity to observe with unprecedented high spatial and spectral resolutions the surface of a cometary nucleus. Moreover, these observations were extended over a long period of time, allowing us to follow the changes occurred on the surface while the heliocentric distance was reducing from 3.6 AU to the frost line at 2.7 AU. During this period of time VIRTIS results point out that 1) single scattering albedo is increasing up to 30$\%$ indicating that the nucleus is becoming brighter; 2) VIS and IR spectral slopes are both decreasing as a consequence of the flattening of the reflectance spectra. Indeed, increasing the water ice abundance in the surface layers will result in an increase of the reflectance as well as in a flattening of the spectrum. The partial removal of the dust layer caused by the start of gaseous activity is the probable cause of the increasing abundance of water ice at the surface; 3)  Active regions localized in the neck/north pole area appear brighter, bluer and with a larger 3.2 $\mu m$ band depth than the rest of the large and small lobes. The increase of the 3.2 $\mu m$ band depth and the shift of its band center toward shorter wavelengths in these areas are also a consequence of the presence of H$_2$O. A multivariate classification based on the Parallel Coordinates method applied to the spectral indicators averaged over geomorphological units shows the presence of three spectral classes identifying smooth/active areas, dust covered areas and depressions while the more numerous consolidated terrains are spread across all these three classes. The difficulty to group together the consolidated terrains could be a consequence of the fact that these units contain many gravitational sinks and topographically low areas \citep{El-Maarry2015} where dust could preferentially accumulate making their spectral behavior more variable.
We plan to follow this investigation in a second paper devoted to a similar study of 67P/CG surface properties and mapping using the data collected during the Rosetta's escort phase, from November 2014 to early May 2015, corresponding to heliocentric distances between 2.93 and 1.71 AU.

\section*{Acknowledgments}
The authors would like to thank the following institutions and agencies, which supported this work: Italian Space Agency (ASI - Italy), Centre National d'Etudes Spatiales (CNES- France), Deutsches Zentrum für Luft- und Raumfahrt (DLR-Germany), National Aeronautic and Space Administration (NASA-USA). VIRTIS was built by a consortium from Italy, France and Germany, under the scientific responsibility of IAPS, Istituto di Astrofisica e Planetologia Spaziali of INAF, Rome (IT), which lead also the scientific operations.  The VIRTIS instrument development for ESA has been funded and managed by ASI, with contributions from Observatoire de Meudon financed by CNES and from DLR. The VIRTIS instrument industrial prime contractor was former Officine Galileo, now Selex ES (Finmeccanica Group) in Campi Bisenzio, Florence, IT. The authors wish to thank the Rosetta Liaison Scientists, the Rosetta Science Ground Segment and the Rosetta Mission Operations Centre for their support in planning the VIRTIS observations. The VIRTIS calibrated data will be available through the ESA's Planetary Science Archive (PSA) web site. This research has made use of NASA's Astrophysics Data System.

\end{document}